%% file: Main.TEX
\documentclass[paper,12pt]{article} 
\usepackage{amsmath}
\usepackage{amsthm}
\usepackage{setspace}
\usepackage{graphicx}

\advance\topmargin by -1.0truein  
\advance\textheight by 2.0truein

\newcommand{\newc}{\newcommand}
 \newc{\1}{{\mathbf 1}}
\newc{\Z}{{\cal Z}}
\newc{\X}{{\cal X}}

\begin{document}

\newpage  
\pagenumbering{arabic}
\setcounter{page}{1}

\title{Statistical Inference on Explained Variation in  High-dimensional Linear Model with Dense Effects}

\author{
   Hua Yun Chen  \\
% Xuelong Wang, Mary Turyk, Maria Argo, and Vectoria Pesky \\
\\
   Division of Epidemiology \& Biostatistics \\
   School of Public Health, University of Illinois at Chicago \\
   1603 West Taylor Street, Chicago, IL 60612 \\
           }
\date{}
\maketitle

\begin{abstract} 
%\baselineskip=2.0\baselineskip
Statistical inference on the explained variation of an outcome by a set of covariates is of particular interest in practice. When the covariates are of moderate to high-dimension and the effects are not sparse, several approaches have been proposed for estimation and inference. One major problem with the existing approaches is that the inference procedures are not robust to the normality assumption on the covariates and the residual errors. In this paper, we propose an estimating equation approach to the estimation and inference on the explained variation in the high-dimensional linear model. Unlike the existing approaches, the proposed approach does not rely on the restrictive normality assumptions for inference. It is shown that the proposed estimator is consistent and asymptotically normally distributed under reasonable conditions. Simulation studies demonstrate better performance of the proposed inference procedure  in comparison with the existing approaches. The proposed approach is applied to studying the variation of glycohemoglobin explained by environmental pollutants in a National Health and Nutrition Examination Survey data set. 
\end{abstract}


\newpage
%\baselineskip=2.0\baselineskip
\input{estequ}

\newpage\input{reference}

%\newpage\input{app}
%\newpage\input{Thm3cond}
\newpage\input{figtab}

% \newpage\input{SuppMater}
%\newpage\input{appclt}

%%%%%%%%%% Merge with supplemental materials %%%%%%%%%%
\clearpage
\begin{center}
   \LARGE Supplemental Material: Statistical Inference on Explained Variation in High-dimensional Linear Model with Dense Effects
\end{center}

\begin{center}
   \author{\large Hua Yun Chen}
\end{center}

\setcounter{section}{0}
\setcounter{table}{0}

\section{Proofs of Theorems}

The proofs of the theorems use the Marchenko-Pastur limit law for the empirical covariance matrix of a random matrix,  we list the law here for easy reference. Let $Z_{n\times p}=(Z_{ij})$ where $Z_{ij}, i=1,\cdots, n; j=1,\cdots,p$ are iid with $E(Z_{ij})=0$ and $E(Z_{ij}^{2})=1$. Let
$
M=\frac{1}{p}Z_{n\times p}Z_{n\times p}^{t}.
$
Let the eigenvalues of $M$ be $\eta_{1},\cdots,\eta_{n}$.  Define the empirical spectral measure as
$$
\mu_{n}(A)=\frac{1}{n}\sum_{i=1}^{n}1_{\{\eta_{i}\in A\}}.
$$
Let $n/p\rightarrow \xi$ as $n,p\rightarrow \infty$, where $\xi\in (0,\infty)$. Then $\mu_{n}\rightarrow \mu$ in distribution, where 
$$
\mu(A)=\left\{\begin{array}{ll}
(1-1/\xi) 1_{\{0\in A\}} +\nu(A)          & \mbox{if } \xi>1,  \\
\nu(A)                                                 &\mbox{if } 0<\xi\leq 1
\end{array}\right.
$$
and the measure $\nu$ has density
$$
d\nu(u)=\frac{1_{\{x\in [a, b]\}}}{2\pi x\xi}\sqrt{(x-a)(b-x)}.
$$
 and $a=(1-\sqrt{\xi})^{2}$ and $b=(1+\sqrt{\xi})^{2}$.

\vspace{0.7cm}

\noindent{\bf Proof of Theorem 1}:  Consider first the case where $n\geq p$. Let the singular value decomposition be
$$
Z_{n\times p}=U_{n\times p}D_{p\times p}V_{p\times p} ^{t},
$$
where $U^{t}U=I_{p\times p}$ and $V^{t}V=VV^{t}=I_{p\times p}$, and $D=\mbox{diag}(d_{1},\cdots,d_{p})$. Let $U^{*}_{n\times(n-p)}$ be the matrix such that
$$
(U,U^{*})^{t}(U,U^{*})=(U,U^{*})(U,U^{*})^{t}=I_{n\times n}.
$$
For $0\leq \lambda<1$, it follows that
$$
I_{n\times n}+\lambda M=(U,U^{*})
\left(\begin{array}{cc}
I_{p\times p}+\lambda D^{2}/p &0 \\
0 & I_{(n-p)\times (n-p)}
\end{array}\right)(U,U^{*})^{t}.
$$
Let $\eta_{k}=d_{k}^{2}/p, k=1,\cdots,p$. We then have
\begin{eqnarray*}
(I_{n\times n}+\lambda M)^{-1}
&=&I_{n\times n}- U\mbox{diag}\left(\frac{\lambda\eta_{1}}{1+\lambda\eta_{1}},\cdots,\frac{\lambda\eta_{p}}{1+\lambda\eta_{p}}\right)U^{t},
\end{eqnarray*}
and
$
W_{\lambda}=U\left(\Delta+I_{p\times p}\right)U^{t}-I_{n\times n},
$
where
$$
\Delta=\mbox{diag}\left(\frac{\eta_{1}-1}{(1+\lambda \eta_{1})^{2}},\cdots,\frac{\eta_{p}-1}{(1+\lambda \eta_{p})^{2}}\right).
$$
It follows that
\begin{eqnarray*}
tr\left\{W_{\lambda}(M-I)\right\}&=&\sum_{k=1}^{p}\frac{(\eta_{k}-1)^{2}}{(1+\lambda \eta_{k})^{2}}+n-p, \\
tr\left\{W_{\lambda}(YY^{t}-I)\right\}&=&Y^{t}(U\Delta U^{t}-U^{*}U^{*t})Y-\sum_{k=1}^{p}\frac{\eta_{k}-1}{(1+\lambda \eta_{k})^{2}}+n-p,
\end{eqnarray*}
Note from $U^{*t}U=0$ that $U^{*t}Z=0$. It follows that
\begin{eqnarray*}
Y^{t}(U\Delta U^{t}-U^{*}U^{*t})Y&=&(Z\alpha)^{t}U\Delta U^{t}Z\alpha+2 (Z\alpha)^{t}U\Delta U^{t}\epsilon +\epsilon^{t}(U\Delta U^{t}-U^{*}U^{*t})\epsilon.
\end{eqnarray*}
It follows from the law of large number for sum of independent variables and for quadratic forms (See the supplemental material for verifying conditions for convergence) that $n^{-1}(Z\alpha)^{t}U\Delta U^{t}\epsilon\rightarrow 0$ in probability and that 
$$
\frac{1}{n}\left[\epsilon^{t}(U\Delta U^{t}-U^{*}U^{*t})\epsilon-E\left\{\epsilon^{t}(U\Delta U^{t}-U^{*}U^{*t})\epsilon\mid Z\right\}\right]\rightarrow 0, 
$$
in probability. These imply that
$$
\frac{1}{n}\left[Y^{t}(U\Delta U^{t}-U^{*}U^{*t})Y-(Z\alpha)^{t}U\Delta U^{t}Z\alpha-E\left\{\epsilon^{t}(U\Delta U^{t}-U^{*}U^{*t})\epsilon\mid Z\right\}\right]\rightarrow 0, 
$$
in probability. Next, $(Z\alpha)^{t}U\Delta U^{t}Z\alpha-tr(Z^{t}U\Delta U^{t}Z/p)r^{2}=0$  when $\alpha=0$. When $\alpha\neq 0$, 
\begin{eqnarray*}
(Z\alpha)^{t}U\Delta U^{t}Z\alpha-tr(Z^{t}U\Delta U^{t}Z/p)r^{2} 
&=&p\left\{\sum_{k=1}^{p}\frac{\eta_{k}(\eta_{k}-1)}{(1+\lambda\eta_{k})^{2}}v_{k}^{2}(u)-\frac{1}{p}\sum_{k=1}^{p}\frac{\eta_{k}(\eta_{k}-1)}{(1+\lambda\eta_{k})^{2}}\right\}r^{2},
\end{eqnarray*}
where 
$
(v_{1}(u),\cdots,v_{p}(u))^{t}=V^{t}u,
$
$u=\alpha/||\alpha||_{2}$, and $||\alpha||_{2}^{2}=\sum_{j=1}^{p}\alpha_{j}^{2}=r^{2}$. Note that
$$
\frac{x(x-1)}{(1+\lambda x)^{2}}
$$
is a bounded function on $x\geq 0$ for $\lambda>0$. From the convergence results on the eigenvectors of the random covariance matrix (Bai et al, 2007, Corollary 2 of Theorem 1), it follows that
$$
\left\{\sum_{k=1}^{p}\frac{\eta_{k}(\eta_{k}-1)}{(1+\lambda\eta_{k})^{2}}v_{k}^{2}(u)-\frac{1}{p}\sum_{k=1}^{p}\frac{\eta_{k}(\eta_{k}-1)}{(1+\lambda\eta_{k})^{2}}\right\}\rightarrow 0, a.s.
$$ 
Note here that the convergence also holds when $\lambda=0$ because of the bounded support of the limited spectral distribution. See also Silverstein (1989).  Furthermore,
\begin{eqnarray*}
E\left\{\epsilon^{t}(U\Delta U^{t}-U^{*}U^{*t})\epsilon\mid Z\right\}&=&(1-r^{2})\left\{\sum_{k=1}^{p}\frac{\eta_{k}-1}{(1+\lambda \eta_{k})^{2}}-(n-p)\right\}, \\
tr(Z^{t}U\Delta U^{t}Z/p)r^{2}&=&r^{2}\sum_{k=1}^{p}\frac{\eta_{k}(\eta_{k}-1)}{(1+\lambda \eta_{k})^{2}},
\end{eqnarray*}
It follows that
\begin{eqnarray*}
\frac{1}{n}\left[Y^{t}(U\Delta U^{t}-U^{*}U^{*t})Y-\sum_{k=1}^{p}\frac{\eta_{k}-1}{(1+\lambda \eta_{k})^{2}}+n-p 
-r^{2}\left\{\sum_{k=1}^{p}\frac{(\eta_{k}-1)^{2}}{(1+\lambda \eta_{k})^{2}}+n-p\right\}\right]&\rightarrow&  0
\end{eqnarray*}
in probability. By the Marchenko-Pastur limit law,
\begin{eqnarray*}
\frac{1}{n}tr\left\{W_{\lambda}(M-I)\right\}&\rightarrow& C=\int_{a}^{b}\frac{(x-1)^{2}}{(1+\lambda x)^{2}}d\nu(x)+(1-1/\xi)1_{\{\xi> 1\}}>0.
\end{eqnarray*}
Since almost sure convergence implies convergence in probability, these put together  imply 
$$
\hat{r}_{\lambda}^{2}\rightarrow r^{2}, 
$$
in probability. It is now straightforward to see that $\hat{\sigma}_{s}^{2}\rightarrow \sigma_{s}^{2}$ and  $\hat{\sigma}_{\epsilon}^{2}\rightarrow \sigma_{\epsilon}^{2}$ in probability.

For $n<p$,  let the singular value decomposition be
$$
Z_{n\times p}=U_{n\times n}D_{n\times n}\{V_{p\times n}\} ^{t},
$$
where $U^{t}U=UU^{t}=I_{n\times n}$ and $V^{t}V=I_{n\times n}$, and $D=\mbox{diag}(d_{1},\cdots,d_{n})$. 
It follows that
\begin{eqnarray*}
W_{\lambda}&=&U\mbox{diag}\left(\frac{\eta_{1}-1}{(1+\lambda\eta_{1})^{2}},\cdots,\frac{\eta_{n}-1}{(1+\lambda \eta_{n})^{2}}\right)U^{t}.
\end{eqnarray*}
We have
\begin{eqnarray*}
\frac{1}{n}tr\left\{W_{\lambda}(M-I)\right\}&=&\frac{1}{n}\sum_{k=1}^{n}\frac{(\eta_{k}-1)^{2}}{(1+\lambda \eta_{k})^{2}}, \\
           &\rightarrow& \int \frac{(x-1)^{2}}{(1+\lambda x)^{2}}dv(x)>0, \\
tr\left\{W_{\lambda}(YY^{t}-I)\right\}&=&Y^{t}U\Delta U^{t}Y-\sum_{k=1}^{n}\frac{\eta_{k}-1}{(1+\lambda \eta_{k})^{2}}.
\end{eqnarray*}
Note further that
$$
Y^{t}U\Delta U^{t}Y=(Z\alpha)^{t}U\Delta U^{t}Z\alpha+2(Z\alpha)^{t}U\Delta U^{t}\epsilon+\epsilon^{t}U\Delta U^{t}\epsilon.
$$
Note from the law of large number for  sum of independent variables and for quadratic forms that $n^{-1}(Z\alpha)^{t}U\Delta U^{t}\epsilon\rightarrow 0$ in probability and
$$
\frac{1}{n}\left[\epsilon^{t}U\Delta U^{t}\epsilon-E\left\{\epsilon^{t}U\Delta U^{t}\epsilon\mid Z\right\}\right]\rightarrow 0  
$$
in probability. It follows that
$$
\frac{1}{n}\left[Y^{t}U\Delta U^{t}Y-(Z\alpha)^{t}U\Delta U^{t}Z\alpha-E\left\{\epsilon^{t}U\Delta U^{t}\epsilon\mid Z\right\}\right]\rightarrow 0
$$
in probability. When $\alpha=0$, $(Z\alpha)^{t}U\Delta U^{t}Z\alpha-tr(Z^{t}U\Delta U^{t}Z/p)r^{2}=0$. When $\alpha\neq 0$, it follows from the results of Bai et al (2007) that
$$
\frac{1}{n}\left\{(Z\alpha)^{t}U\Delta U^{t}Z\alpha-tr(Z^{t}U\Delta U^{t}Z/p)r^{2}\right\}\rightarrow 0, a.s.
$$
It follows from 
\begin{eqnarray*}
E\left\{\epsilon^{t}U\Delta U^{t}\epsilon\mid Z\right\}&=&(1-r^{2})\sum_{k=1}^{n}\frac{\eta_{k}-1}{(1+\lambda \eta_{k})^{2}}, \\
tr(Z^{t}U\Delta U^{t}Z/p)\sigma_{s}^{2}&=&r^{2}\sum_{k=1}^{n}\frac{\eta_{k}(\eta_{k}-1)}{(1+\lambda \eta_{k})^{2}},
\end{eqnarray*}
that
\begin{eqnarray*}
\frac{1}{n}\left[Y^{t}U\Delta U^{t}Y-\sum_{k=1}^{n}\frac{\eta_{k}-1}{(1-\lambda+\lambda \eta_{k})^{2}}-r^{2}\sum_{k=1}^{n}\frac{(\eta_{k}-1)^{2}}{(1+\lambda \eta_{k})^{2}}\right]&\rightarrow&  0
\end{eqnarray*}
in probability. Hence,
$
\hat{r}_{\lambda}^{2}\rightarrow r^{2}
$ 
in probability, and thus  $\hat{\sigma}_{s}^{2}\rightarrow \sigma_{s}^{2}$ and  $\hat{\sigma}_{\epsilon}^{2}\rightarrow \sigma_{\epsilon}^{2}$ in probability. 

%%%%%%%%%%%%%%%%%%%
%\begin{comment}
%%%%%%%%%%%%%%%%%%%
\noindent{\bf Proof of Theorem 1}: For $n\geq p$, note from $U^{*t}U=0$ that $U^{*t}Z=0$. It follows that
\begin{eqnarray*}
Y^{t}(U\Delta U^{t}-U^{*}U^{*t})Y&=&(Z\alpha)^{t}U\Delta U^{t}Z\alpha+2 (Z\alpha)^{t}U\Delta U^{t}\epsilon +\epsilon^{t}(U\Delta U^{t}-U^{*}U^{*t})\epsilon.
\end{eqnarray*}
It follows from the law of large number for sum of independent variables and for quadratic forms (See the supplemental material for verifying conditions for convergence) that $n^{-1}(Z\alpha)^{t}U\Delta U^{t}\epsilon\rightarrow 0$ in probability and that 
$$
\frac{1}{n}\left[\epsilon^{t}(U\Delta U^{t}-U^{*}U^{*t})\epsilon-E\left\{\epsilon^{t}(U\Delta U^{t}-U^{*}U^{*t})\epsilon\mid Z\right\}\right]\rightarrow 0, 
$$
in probability. These imply that
$$
\frac{1}{n}\left[Y^{t}(U\Delta U^{t}-U^{*}U^{*t})Y-(Z\alpha)^{t}U\Delta U^{t}Z\alpha-E\left\{\epsilon^{t}(U\Delta U^{t}-U^{*}U^{*t})\epsilon\mid Z\right\}\right]\rightarrow 0, 
$$
in probability. Next, $(Z\alpha)^{t}U\Delta U^{t}Z\alpha-tr(Z^{t}U\Delta U^{t}Z/p)r^{2}=0$  when $\alpha=0$. When $\alpha\neq 0$, 
\begin{eqnarray*}
(Z\alpha)^{t}U\Delta U^{t}Z\alpha-tr(Z^{t}U\Delta U^{t}Z/p)r^{2} 
&=&p\left\{\sum_{k=1}^{p}\frac{\eta_{k}(\eta_{k}-1)}{(1+\lambda\eta_{k})^{2}}v_{k}^{2}(u)-\frac{1}{p}\sum_{k=1}^{p}\frac{\eta_{k}(\eta_{k}-1)}{(1+\lambda\eta_{k})^{2}}\right\}r^{2},
\end{eqnarray*}
where 
$
(v_{1}(u),\cdots,v_{p}(u))^{t}=V^{t}u,
$
$u=\alpha/||\alpha||_{2}$, and $||\alpha||_{2}^{2}=\sum_{j=1}^{p}\alpha_{j}^{2}=r^{2}$. Note that
$$
\frac{x(x-1)}{(1+\lambda x)^{2}}
$$
is a bounded function on $x\geq 0$ for $\lambda>0$. From the convergence results on the eigenvectors of the random covariance matrix (Bai et al, 2007, Corollary 2 of Theorem 1), it follows that
$$
\left\{\sum_{k=1}^{p}\frac{\eta_{k}(\eta_{k}-1)}{(1+\lambda\eta_{k})^{2}}v_{k}^{2}(u)-\frac{1}{p}\sum_{k=1}^{p}\frac{\eta_{k}(\eta_{k}-1)}{(1+\lambda\eta_{k})^{2}}\right\}\rightarrow 0, a.s.
$$ 
Note here that the convergence also holds when $\lambda=0$ because of the bounded support of the limited spectral distribution. See also Silverstein (1989).  Furthermore,
\begin{eqnarray*}
E\left\{\epsilon^{t}(U\Delta U^{t}-U^{*}U^{*t})\epsilon\mid Z\right\}&=&(1-r^{2})\left\{\sum_{k=1}^{p}\frac{\eta_{k}-1}{(1+\lambda \eta_{k})^{2}}-(n-p)\right\}, \\
tr(Z^{t}U\Delta U^{t}Z/p)r^{2}&=&r^{2}\sum_{k=1}^{p}\frac{\eta_{k}(\eta_{k}-1)}{(1+\lambda \eta_{k})^{2}},
\end{eqnarray*}
It follows that
\begin{eqnarray*}
\frac{1}{n}\left[Y^{t}(U\Delta U^{t}-U^{*}U^{*t})Y-\sum_{k=1}^{p}\frac{\eta_{k}-1}{(1+\lambda \eta_{k})^{2}}+n-p 
-r^{2}\left\{\sum_{k=1}^{p}\frac{(\eta_{k}-1)^{2}}{(1+\lambda \eta_{k})^{2}}+n-p\right\}\right]&\rightarrow&  0
\end{eqnarray*}
in probability. By the Marchenko-Pastur limit law,
\begin{eqnarray*}
\frac{1}{n}tr\left\{W(M-I)\right\}&\rightarrow& C=\int_{a}^{b}\frac{(x-1)^{2}}{(1+\lambda x)^{2}}d\nu(x)+(1-1/\xi)1_{\{\xi\geq 1\}}>0.
\end{eqnarray*}
Since almost sure convergence implies convergence in probability, these put together  imply 
$$
\hat{r}^{2}\rightarrow r^{2}, 
$$
in probability. It is now straightforward to see that $\hat{\sigma}_{s}^{2}\rightarrow \sigma_{s}^{2}$ and  $\hat{\sigma}_{\epsilon}^{2}\rightarrow \sigma_{\epsilon}^{2}$ in probability.

For the case $n<p$,  note that
$$
Y^{t}U\Delta U^{t}Y=(Z\alpha)^{t}U\Delta U^{t}Z\alpha+2(Z\alpha)^{t}U\Delta U^{t}\epsilon+\epsilon^{t}U\Delta U^{t}\epsilon.
$$
Note from the law of large number for  sum of independent variables and for quadratic forms that $n^{-1}(Z\alpha)^{t}U\Delta U^{t}\epsilon\rightarrow 0$ in probability and
$$
\frac{1}{n}\left[\epsilon^{t}U\Delta U^{t}\epsilon-E\left\{\epsilon^{t}U\Delta U^{t}\epsilon\mid Z\right\}\right]\rightarrow 0  
$$
in probability. It follows that
$$
\frac{1}{n}\left[Y^{t}U\Delta U^{t}Y-(Z\alpha)^{t}U\Delta U^{t}Z\alpha-E\left\{\epsilon^{t}U\Delta U^{t}\epsilon\mid Z\right\}\right]\rightarrow 0
$$
in probability. When $\alpha=0$, $(Z\alpha)^{t}U\Delta U^{t}Z\alpha-tr(Z^{t}U\Delta U^{t}Z/p)r^{2}=0$. When $\alpha\neq 0$, it follows from the results of Bai et al (2007) that
$$
\frac{1}{n}\left\{(Z\alpha)^{t}U\Delta U^{t}Z\alpha-tr(Z^{t}U\Delta U^{t}Z/p)r^{2}\right\}\rightarrow 0, a.s.
$$
It follows from 
\begin{eqnarray*}
E\left\{\epsilon^{t}U\Delta U^{t}\epsilon\mid Z\right\}&=&(1-r^{2})\sum_{k=1}^{n}\frac{\eta_{k}-1}{(1+\lambda \eta_{k})^{2}}, \\
tr(Z^{t}U\Delta U^{t}Z/p)\sigma_{s}^{2}&=&r^{2}\sum_{k=1}^{n}\frac{\eta_{k}(\eta_{k}-1)}{(1+\lambda \eta_{k})^{2}},
\end{eqnarray*}
that
\begin{eqnarray*}
\frac{1}{n}\left[Y^{t}U\Delta U^{t}Y-\sum_{k=1}^{n}\frac{\eta_{k}-1}{(1-\lambda+\lambda \eta_{k})^{2}}-r^{2}\sum_{k=1}^{n}\frac{(\eta_{k}-1)^{2}}{(1+\lambda \eta_{k})^{2}}\right]&\rightarrow&  0
\end{eqnarray*}
in probability. Hence,
$
\hat{r}^{2}\rightarrow r^{2}
$ 
in probability, and thus  $\hat{\sigma}_{s}^{2}\rightarrow \sigma_{s}^{2}$ and  $\hat{\sigma}_{\epsilon}^{2}\rightarrow \sigma_{\epsilon}^{2}$ in probability.
%%%%%%%%%%%%% 
%\end{comment}
%%%%%%%%%%%%%

\vspace{0.7cm}

\noindent {\bf Proof of Theorem 2}: Consider the case  $\sigma_{s}^{2}+\sigma_{\epsilon}^{2}=1$ and $r^{2}=\sigma_{s}^{2}$. It follows that
$$
\hat{r}^{2}-r^{2}=\frac{tr\left\{W(YY^{t}-M\sigma_{s}^{2}-I\sigma_{\epsilon}^{2})\right\}}
{tr\left\{W(M-I)\right\}},
$$
Since 
$$
YY^{t}-M\sigma_{s}^{2}-I\sigma_{\epsilon}^{2}=Z\alpha(Z\alpha)^{t}-M\sigma_{s}^{2}+Z\alpha\epsilon^{t}+\epsilon(Z\alpha)^{t}+\epsilon\epsilon^{t}-I\sigma_{\epsilon}^{2},
$$
it follows that
\begin{eqnarray*}
\frac{1}{\sqrt{n}}tr\{W(\tilde{Y}\tilde{Y}'-M\sigma_{s}^{2}-I\sigma_{\epsilon}^{2})\}&=&\frac{1}{\sqrt{n}}\left\{\alpha^{t}Z^{t}WZ\alpha-\frac{1}{p}tr(Z^{t}WZ)\sigma_{s}^{2}\right\}+\frac{2}{\sqrt{n}}\epsilon^{t}WZ\alpha \\
&&+\frac{1}{\sqrt{n}}\left\{\epsilon^{t}W\epsilon-tr(W)\sigma_{\epsilon}^{2}\right\},
\end{eqnarray*}
When $\alpha=0$, 
\begin{eqnarray*}
\frac{1}{\sqrt{n}}tr\{W(\tilde{Y}\tilde{Y}'-M\sigma_{s}^{2}-I\sigma_{\epsilon}^{2})\}&=&\frac{1}{\sqrt{n}}\left\{\epsilon^{t}W\epsilon-tr(W)\sigma_{\epsilon}^{2}\right\} \\
&=&\frac{1}{\sqrt{n}}\sum_{i=1}^{n}\sum_{j\neq i}W_{ij}\epsilon_{i}\epsilon_{j}+\frac{1}{\sqrt{n}}\sum_{i=1}^{n}W_{ii}(\epsilon_{i}^{2}-\sigma_{\epsilon}^{2}).
\end{eqnarray*}
Conditional on $Z$, it follows from the central limit theorems for independent variables and for the quadratic forms (de Jong, 1987 and Heyde and Brown, 1970.  See supplemental material for the condition verification) that
$$
\frac{1}{\sqrt{n}}\left(\begin{array}{c}
\sum_{i=1}^{n}\sum_{j\neq i}W_{ij}\epsilon_{i}\epsilon_{j}, \\
\sum_{i=1}^{n}W_{ii}(\epsilon_{i}^{2}-\sigma_{\epsilon}^{2})
\end{array}\right)
\rightarrow 
N\left(\left(\begin{array}{c}0\\0\end{array}\right), 
\left(\begin{array}{cc}
2\sigma_{\epsilon}^{4}(S-T) &0 \\
0 & TE(\epsilon^{2}-\sigma_{\epsilon}^{2})^{2}
\end{array}\right)\right),
$$
where
\begin{eqnarray*}
S&=&\lim _{n\rightarrow\infty}\frac{1}{n}tr(W^{2})  \\ 
T&=&\lim _{n\rightarrow\infty}\frac{1}{n}\sum_{i=1}^{n}W_{ii}^{2}.
\end{eqnarray*}
It follows that
$$
\sqrt{n}(\hat{r}^{2}-r^{2})\rightarrow N\left(0, \frac{2\sigma_{\epsilon}^{4}(S-T)+TE(\epsilon^{2}-\sigma_{\epsilon}^{2})^{2}}{C^{2}}\right).
$$ 

For $\alpha\neq 0$,  
$$
\frac{1}{\sqrt{n}}\left\{\alpha^{t}Z^{t}WZ\alpha-\frac{1}{p}tr(Z^{t}WZ)\sigma_{s}^{2}\right\}= \frac{p}{\sqrt{n}}\left\{\sum_{k=1}^{p}\frac{\eta_{k}(\eta_{k}-1)}{(1+\lambda\eta_{k})^{2}}v_{k}^{2}(u)-\frac{1}{p}\sum_{k=1}^{p}\frac{\eta_{k}(\eta_{k}-1)}{(1+\lambda\eta_{k})^{2}}\right\}\sigma_{s}^{2} 
$$
when $n\geq p$, and 
$$
\frac{1}{\sqrt{n}}\left\{\alpha^{t}Z^{t}WZ\alpha-\frac{1}{p}tr(Z^{t}WZ)\sigma_{s}^{2}\right\}= \frac{p}{\sqrt{n}}\left\{\sum_{k=1}^{n}\frac{\eta_{k}(\eta_{k}-1)}{(1+\lambda\eta_{k})^{2}}v_{k}^{2}(u)-\frac{1}{p}\sum_{k=1}^{n}\frac{\eta_{k}(\eta_{k}-1)}{(1+\lambda\eta_{k})^{2}}\right\}\sigma_{s}^{2}
$$
when  $n<p$.  It follows by applying Thorem 3 of Bai et al (2007) or Theorem 1.3 of Pan and Zhou (2008) that
$$
 \sqrt{\frac{p}{2}}\left\{\sum_{k=1}^{p\wedge n}\frac{\eta_{k}(\eta_{k}-1)}{(1+\lambda\eta_{k})^{2}}v_{k}^{2}(u)-\frac{1}{p}\sum_{k=1}^{p\wedge n}\frac{\eta_{k}(\eta_{k}-1)}{(1+\lambda\eta_{k})^{2}}\right\}\rightarrow N(0, \tau^{2}),
$$
where
$$
\tau^{2}=\int_{a}^{b} \left\{\frac{x(x-1)}{(1+\lambda x)^{2}}\right\}^{2}d\nu(x)-\left\{\int_{a}^{b} \frac{x(x-1)}{(1+\lambda x)^{2}}d\nu(x)\right\}^{2},
$$
and $\nu(x)$ is the limiting spectral measure in the Marchenko-Pastur law. It follows that
$$
\frac{1}{\sqrt{n}}\left\{\alpha^{t}Z^{t}WZ\alpha-\frac{1}{p}tr(Z^{t}WZ)\sigma_{s}^{2}\right\}\rightarrow N(0, 2\tau^{2}\sigma_{s}^{4}/\xi),
$$
Conditional on $Z$, it follows from the central limit theorems for independent variables and for the quadratic forms (Heyde and Brown, 1970.  See supplemental materials for the condition verification) that
$$
\frac{1}{\sqrt{n}}\left(\begin{array}{c}
\sum_{i=1}^{n}(WZ\alpha)_{i}\epsilon_{i} \\
\sum_{i=1}^{n}\sum_{j\neq i}W_{ij}\epsilon_{i}\epsilon_{j} \\
\sum_{i=1}^{n}W_{ii}(\epsilon_{i}^{2}-\sigma_{\epsilon}^{2})
\end{array}\right)
\rightarrow 
N\left(\left(\begin{array}{c}0\\0\\0\end{array}\right), 
\left(\begin{array}{ccc}
 \sigma_{\epsilon}^{2}\sigma_{s}^{2}S_{1} &   0 & T_{1}E\epsilon^{3} \\
0&2\sigma_{\epsilon}^{4}(S-T) &0 \\
T_{1}E\epsilon^{3}&0 & TE(\epsilon^{2}-\sigma_{\epsilon}^{2})^{2}
\end{array}\right)\right),
$$
where 
\begin{eqnarray*}
S_{1}&=&\frac{1}{n}\lim_{n\rightarrow \infty}tr(W^{t}WM), \\
T_{1}&=&\lim_{n\rightarrow\infty}\frac{1}{n}\sum_{i=1}^{n}(WZ\alpha)_{i}W_{ii},
\end{eqnarray*}
Together, we have
$
\sqrt{n}(\hat{r}^{2}-r^{2})\rightarrow N(0, v^{2}),
$ 
where
$$
v^{2}=\frac{2\tau^{2}\sigma_{s}^{4}/\xi+4\sigma_{\epsilon}^{2}\sigma_{s}^{2}S_{1}+2\sigma_{\epsilon}^{4}(S-T) +TE(\epsilon^{2}-\sigma_{\epsilon}^{2})^{2}+2T_{1}E(\epsilon^{3})}
{C^{2}}
$$

\vspace{0.3cm}

\noindent {\bf Proof of Theorem 3}: Note that $W_{*}=I-Z(Z^{t}Z)^{-1}Z^{t}$ and
\begin{eqnarray*}
\hat{r}_{*}^{2}&=&1-\frac{1}{n-p}Y^{t}\left\{I-Z(Z^{t}Z)^{-1}Z^{t}\right\}Y. \\
&=&1-\frac{1}{n-p}\epsilon^{t}\left\{I-Z(Z^{t}Z)^{-1}Z^{t}\right\}\epsilon. 
\end{eqnarray*}
 Since for fixed symmetric matrix $A=(a_{ij})_{n\times n}$,
\begin{equation}
E(\epsilon^{t} A\epsilon)^{2}=\sum_{i=1}^{n}a_{ii}^{2}(E\epsilon^{4}-3\sigma_{\epsilon}^{4})+\left[2tr(A^{2})+\{tr(A)\}^{2}\right]\sigma_{\epsilon}^{4}.
\label{QM}
\end{equation}
It follows that, 
\begin{eqnarray*}
E\left\{|\hat{r}_{*}^{2}-r^{2}|^{2}\mid Z\right\}&=&\frac{1}{(n-p)^{2}}\sum_{i=1}^{n}a_{ii}^{2}\left(E\epsilon^{4}-3\sigma_{\epsilon}^{4}\right)+\frac{2}{n-p}\sigma_{\epsilon}^{4} \\
&\leq&\frac{1}{n-p}\left(E\epsilon^{4}-\sigma_{\epsilon}^{4}\right)\rightarrow 0,
\end{eqnarray*}
because $n-p=n(1-1/\xi)\rightarrow \infty$. It follows that $\hat{r}_{*}^{2}\rightarrow r^{2}$ in probability.

Conditional on $Z$, it follows from the central limit theorems for independent variables and for the quadratic forms (de Jong, 1987 and Heyde and Brown, 1970) that
$$
\frac{1}{\sqrt{n}}\left(\begin{array}{c}
\sum_{i=1}^{n}\sum_{j\neq i}W_{* ij}\epsilon_{i}\epsilon_{j}, \\
\sum_{i=1}^{n}W_{* ii}(\epsilon_{i}^{2}-\sigma_{\epsilon}^{2})
\end{array}\right)
\rightarrow 
N\left(\left(\begin{array}{c}0\\0\end{array}\right), 
\left(\begin{array}{cc}
2\sigma_{\epsilon}^{4}(S_{*}-T_{*}) &0 \\
0 & T_{*}E(\epsilon^{2}-\sigma_{\epsilon}^{2})^{2}
\end{array}\right)\right),
$$
where
\begin{eqnarray*}
S_{*}&=&\lim _{n\rightarrow\infty}\frac{1}{n}tr(W_{*}^{2}) =1-1/\xi, \\ 
T_{*}&=&\lim _{n\rightarrow\infty}\frac{1}{n}\sum_{i=1}^{n}W_{* ii}^{2}=(1-1/\xi)^{2}.
\end{eqnarray*}
Conditions for the central limit theorem to hold can be verified similarly as in Theorem 2 except that it is simpler because no $\alpha$ is involved.  
It follows that
$$
\sqrt{n}(\hat{r}_{*}^{2}-r^{2})\rightarrow N(0, v_{*}^{2}) ,
$$ 
where
$$
v_{*}^{2}=\frac{2\sigma_{\epsilon}^{4}}{(\xi-1)}+E(\epsilon^{2}-\sigma_{\epsilon}^{2})^{2}.
$$

\section{Variance estimates for the explained variation estimators}

\vspace{0.3cm}

When $\alpha=0$, the numerator in the $v_{\lambda}^{2}$ expression can be approximated by
$$
\frac{2\sigma_{\epsilon}^{4}}{n}tr(W_{\lambda}^{t}W_{\lambda})+\frac{1}{n}\sum_{i=1}^{n}W_{\lambda ii}^{2}\left(E\epsilon^{4}-3\sigma_{\epsilon}^{4}\right),
$$
which can be consistently estimated by
$$
\frac{2}{n}\left\{tr(W_{\lambda}^{t}W_{\lambda})-\sum_{i=1}^{n}W_{\lambda ii}^{2}\right\}+\frac{1}{n}\sum_{i=1}^{n}W_{\lambda ii}^{2}(Y_{i}^{2}-1)^{2}.
$$

In general, to estimate the variance $v_{\lambda}^{2}$, note that $C$ can be consistently estimated by $tr\{W_{\lambda}(M-I)\}/n$. The first term in the numerator can be consistently estimated by $2\hat{r}^{4}\hat{\tau}^{2}p/n$, where
$$
\hat{\tau}^{2}=\frac{1}{p}\sum_{k=1}^{p\wedge n}\left\{\frac{\eta_{k}(\eta_{k}-1)}{(1+\lambda\eta_{k})^{2}}\right\}^{2}-\left\{\frac{1}{p}\sum_{k=1}^{p\wedge n}\frac{\eta_{k}(\eta_{k}-1)}{(1+\lambda\eta_{k})^{2}}\right\}^{2}.
$$
The second term in the numerator can be consistently estimated by $4\hat{r}^{2}(1-\hat{r}^{2})tr(W_{\lambda}^{t}W_{\lambda}M)/n$. The first part of the third term ($2\sigma_{\epsilon}^{2}S$) in the numerator can be consistently estimated by $2(1-\hat{r}^{2})^{2}tr(W_{\lambda}^{t}W_{\lambda})/n$. 
When $\epsilon$ follows the normal distribution, the rest terms in the numerator, i.e., 
$$
T(E\epsilon^{4}-3\sigma_{\epsilon}^{4})+2T_{1}E\epsilon^{3}=0.
$$
As a result,
$$
\tilde{v}_{\lambda}^{2}=\frac{2\hat{r}^{4}\hat{\tau}^{2}p/n+4\hat{r}^{2}(1-\hat{r}^{2})tr(W_{\lambda}^{t}W_{\lambda}M)/n+2(1-\hat{r}^{2})^{2}tr(W_{\lambda}^{t}W_{\lambda})/n}{[tr\{W_{\lambda}(M-I)\}/n]^{2}},
$$
is a consistent estimate of the variance under the normal random error. 
  
For non-zero $\alpha$ and possibly non-normal random error in general, note that
$$
Y_{i}^{2}-1-(M_{ii}-1)\sigma_{s}^{2}=\epsilon_{i}^{2}-\sigma_{\epsilon}^{2}+2\epsilon_{i}z_{i}^{t}\alpha+(z_{i}^{t}\alpha)^{2}-M_{ii} \sigma_{s}^{2}.
$$
It follows that
\begin{eqnarray*}
\sum_{i=1}^{n}E\{Y_{i}^{2}-1-(M_{ii}-1)\sigma_{s}^{2}\}^{2}&=&\sum_{i=1}^{n}E\{\epsilon_{i}^{2}-\sigma_{\epsilon}^{2}\}^{2}+4n\sigma_{s}^{2}\sigma_{\epsilon}^{2}
+\sum_{i=1}^{n}E\{(z_{i}^{t}\alpha)^{2}-M_{ii} \sigma_{s}^{2}\}^{2}.
\end{eqnarray*}
Note that 
$$
\sum_{i=1}^{n}E\{(z_{i}^{t}\alpha)^{2}-M_{ii} \sigma_{s}^{2}\}^{2}=\sum_{i=1}^{n}E\left\{M_{ii}-(z_{i}^{t}u)^{2}\right\}^{2}\sigma_{s}^{4},
$$
where $||u||_{2}=1$. An estimated upper bound for $E(\epsilon^{2}-\sigma_{\epsilon}^{2})^{2}$ is
$$
\widetilde{var(\epsilon^{2})}=
\frac{1}{n}\sum_{i=1}^{n}\{Y_{i}^{2}-1-(M_{ii}-1)\hat{r}^{2}\}^{2} -4\hat{r}^{2}(1-\hat{r}^{2})-g\hat{r}^{4},
$$ 
where
$$
g=\min _{||u||_{2}=1}\frac{1}{n}\sum_{i=1}^{n}E\{(z_{i}^{t}u)^{2}-M_{ii}\}^{2}.
$$
For simplicity of calculation, we may take $g=0$. Recall that $M_{ii}=p^{-1}\sum_{j=1}^{p}Z_{ij}^{2}$. Direct calculation shows that
$$
E\left\{M_{ii}-(z_{i}^{t}u)^{2}\right\}^{2}=\left\{2+\sum_{j=1}^{p}u_{j}^{4}(EZ_{ij}^{4}-3)+p^{-1}\sum_{j=1}^{p}(p^{-1}-2u_{j}^{2})E(Z_{ij}^{4}-1)\right\}.
$$
The last term on the right-hand side is negligible relative to other terms when $p$ is large. For $EZ_{ij}^{4}=3$ or $\max_{i}|\alpha_{i}|\rightarrow 0$,  the second term is $0$. it follows that $g\approx 2$.  When $E\epsilon^{3}=0$, a consistent estimate of the asymptotic variance for $\sqrt{n}(\hat{r}^{2}-r^{2})$ is
$\hat{v}_{\lambda}^{2}=\tilde{v}_{\lambda}^{2}+\widehat{va}$, where
$$
\widehat{va}=\frac{\frac{1}{n}\sum_{i=1}^{n}\{Y_{i}^{2}-1-(M_{ii}-1)\hat{r}^{2}\}^{2} -4\hat{r}^{2}(1-\hat{r}^{2})-2\hat{r}^{4}}{[tr\{W(M-I)\}/n]^{2}}\frac{1}{n}\sum_{i=1}^{n}W_{\lambda ii}^{2}.
$$
Because $\widehat{va}$ is not guaranteed to be postive while the estimand is positive, we propose to instead use 
$$
\hat{v}_{\lambda}^{2}=\tilde{v}_{\lambda}^{2}-\frac{2(1-\hat{r}^{2})^{2}\frac{1}{n}\sum_{i=1}^{n}W_{\lambda ii}^{2}}{[tr\{W(M-I)\}/n]^{2}}+\widehat{va}1_{\{\widehat{va}\geq 0\}}
$$ 
Without assuming $E\epsilon^{3}=0$, 
$$
T_{1}\leq \sigma_{s}\sqrt{S_{1}T} \mbox{ and } E\epsilon^{3}\leq \left\{E(\epsilon^{2}-\sigma_{\epsilon}^{2})^{2}E\epsilon^{2}\right\}^{1/2}.
$$
A variance bound for $v_{\lambda}^{2}$ is
$$
\hat{v}_{\lambda}^{2}+\frac{\hat{r}\sqrt{S_{1}T}\left\{\widehat{va}1_{\{\widehat{va}\geq 0\}}(1-\hat{r}^{2})\right\}^{1/2}}{[tr\{W(M-I)\}/n]^{2}}.
$$

When $\epsilon$ follows the normal distribution, $E\epsilon^{4}=3\sigma_{\epsilon}^{4}$. A consistent estimator of $v_{*}^{2}$ can be obtained as
$$
\hat{v}_{*}^{2}=\frac{2n(1-\hat{r}_{*}^{2})^{2}}{n-p}
$$
When the normality assumption on $\epsilon$ may be voilated, a consistent estimate for $v_{*}^{2}$ is
$$
\hat{v}_{*}^{2}=\frac{2p(1-\hat{r}_{*}^{2})^{2}}{n-p}+\widehat{Var(\epsilon^{2})},
$$
where $\widehat{Var(\epsilon^{2})}$ can be obtained as follows. Recall that the singular value decomposition has
$$
Z_{n\times p}=U_{n\times p}D_{p\times p}V_{p\times p} ^{t},
$$
where $U^{t}U=I_{p\times p}$ and $V^{t}V=VV^{t}=I_{p\times p}$, and $D=\mbox{diag}(d_{1},\cdots,d_{p})$. Furthemore, $U^{*}_{n\times(n-p)}$ is the matrix such that
$$
(U,U^{*})^{t}(U,U^{*})=(U,U^{*})(U,U^{*})^{t}=I_{n\times n}.
$$
It follows from $U^{*t}Z=0$ that
$
U^{*t}Y=U^{*t}\epsilon.
$
It further follows that
$$
E\left\{\sum_{j=1}^{n-p}\left(U_{j}^{*t}\epsilon\right)^{4}\right\}=\left(E\epsilon^{4}-3\sigma_{\epsilon}^{4}\right)\sum_{j=1}^{n-p}\sum_{k=1}^{n}U_{jk}^{*4}+3(n-p)\sigma_{\epsilon}^{4},
$$
where $U^{*t}=(U_{1}^{*t},\cdots,U_{n-p}^{*t})$. This means that $var(\epsilon^{2})$ can be estimated by
$$
\widetilde{var(\epsilon^{2})}=\frac{\sum_{j=1}^{n-p}\left(U_{j}^{*t}Y\right)^{4}-3(n-p)(1-\hat{r}_{*}^{2})^{2}}{\sum_{j=1}^{n-p}\sum_{k=1}^{n}U_{jk}^{*4}}+2(1-\hat{r}^{2})^{2}.
$$
where $U_{j}^{*}=(U_{j1}^{*},\cdots,U_{jn}^{*})$. Since $Var(\epsilon^{2})\geq 0$, we set
$$
\widehat{var(\epsilon^{2})}=\max\left\{\widetilde{var(\epsilon^{2})}, 0\right\}.
$$

\section{Verifying conditions for the consistency proof}

\vspace{0.5cm}

\begin{enumerate}
\item 
$
\frac{1}{n}(Z\alpha)^{t}U\Delta U^{t}\epsilon\rightarrow 0
$
in probability.

\noindent Proof: for any $x>0$, 
\begin{eqnarray*}
P\left\{|\frac{1}{n}(Z\alpha)^{t}U\Delta U^{t}\epsilon|>x\mid Z\right\}\leq \frac{\sigma_{\epsilon}^{2}}{n^{2}x^{2}}\alpha^{t}Z^{t}W^{t}WZ\alpha.
\end{eqnarray*}
From Bai et al (2007), 
$
\frac{1}{n}\alpha^{t}Z^{t}W^{t}WZ\alpha
$ 
converges almost sure to a constant and the sequence is bounded. Hence,
\begin{eqnarray*}
P\left\{|\frac{1}{n}(Z\alpha)^{t}U\Delta U^{t}\epsilon|>x\right\}\leq \frac{\sigma_{\epsilon}^{2}}{n^{2}x^{2}}E\{\alpha^{t}Z^{t}W^{t}WZ\alpha\}\rightarrow 0.
\end{eqnarray*}

\item
$
\frac{1}{n}\left[\epsilon^{t}U\Delta U^{t}\epsilon-E\left\{\epsilon^{t}U\Delta U^{t}\epsilon\mid Z\right\}\right]\rightarrow 0
$
in probability.

\noindent Proof: For any $x>0$, 
\begin{eqnarray*}
&&\hspace{-1cm}P\left\{|\frac{1}{n}\left[\epsilon^{t}U\Delta U^{t}\epsilon-E\left\{\epsilon^{t}U\Delta U^{t}\epsilon\mid Z\right\}\right]|>x\mid Z\right\} \\
&\leq &\frac{1}{n^{2}x^{2}}\left[\sum_{i=1}^{n}\{(U\Delta U^{t})_{ii}\}^{2}\{E(\epsilon^{2}-\sigma_{\epsilon}^{2})^{2}-2\sigma_{\epsilon}^{4}\}+2tr(\Delta^{2})\sigma_{\epsilon}^{4}\right]
\end{eqnarray*}
Since
$$
\frac{1}{n}\sum_{i=1}^{n}\{(U\Delta U^{t})_{ii}\}^{2}\leq \frac{1}{n}tr(\Delta^{2})
$$
converges to a constant by the Marchenko-Pastur law. From the boundedness of the sequence, it follows that
$$
P\left\{|\frac{1}{n}\left[\epsilon^{t}U\Delta U^{t}\epsilon-E\left\{\epsilon^{t}U\Delta U^{t}\epsilon\mid Z\right\}\right]|>x\right\}\rightarrow 0.
$$

\item 
 $
\frac{1}{n}\left[\epsilon^{t}U^{*} U^{*t}\epsilon-E\left\{\epsilon^{t}U^{*} U^{*t}\epsilon\mid Z\right\}\right]\rightarrow 0
$
in probability.

\noindent Proof: For any $x>0$, 
\begin{eqnarray*}
&&\hspace{-1cm}P\left\{|\frac{1}{n}\left[\epsilon^{t}U^{*}U^{*t}\epsilon-E\left\{\epsilon^{t}U^{*}U^{*t}\epsilon\mid Z\right\}\right]|>x\mid Z\right\} \\
&\leq &\frac{1}{n^{2}x^{2}}\left[\sum_{i=1}^{n}\{(U^{*} U^{*t})_{ii}\}^{2}\{E(\epsilon^{2}-\sigma_{\epsilon}^{2})^{2}-2\sigma_{\epsilon}^{4}\} +2tr(U^{*}U^{*t}U^{*}U^{*t})\sigma_{\epsilon}^{4}\right] \\
&\leq& \frac{n-p}{n^{2}x^{2}}E(\epsilon^{2}-\sigma_{\epsilon}^{2})^{2}.
\end{eqnarray*}
It follows that
$$
P\left\{|\frac{1}{n}\left[\epsilon^{t}U\Delta U^{t}\epsilon-E\left\{\epsilon^{t}U\Delta U^{t}\epsilon\mid Z\right\}\right]|>x\right\}\rightarrow 0.
$$

\end{enumerate}

\section{Verifying conditions for convergence in distribution}
\begin{eqnarray*}
&&\hspace{-2cm}\frac{1}{\sqrt{n}}\sum_{i=1}^{n}\left\{(WZ\alpha)_{i}\epsilon_{i}+\sum_{j=1}^{n}W_{ij}\epsilon_{i}\epsilon_{j}-W_{ii}\sigma_{\epsilon}^{2}\right\}  \\
&=&\frac{1}{\sqrt{n}}\sum_{i=1}^{n} \left[\left\{(WZ\alpha)_{i} +2\sum_{j=1}^{i-1}W_{ij}\epsilon_{j}\right\}\epsilon_{i} +W_{ii}(\epsilon_{i}^{2}-\sigma_{\epsilon}^{2})\right].
\end{eqnarray*}
Let ${\cal F}_{ni}=\sigma\{\epsilon_{1},\cdots,\epsilon_{i}, Z\}$. $\{R_{ni}, {\cal F}_{ni}\}$ is a martingale difference sequence, where
$$
R_{ni}=\left\{(WZ\alpha)_{i} +2\sum_{j=1}^{i-1}W_{ij}\epsilon_{j}\right\}\epsilon_{i} +W_{ii}(\epsilon_{i}^{2}-\sigma_{\epsilon}^{2}).
$$
Let $\sigma_{ni}^{2}=E(R_{ni}^{2}\mid {\cal F}_{n(i-1)})$ and $s_{n}=\sum_{i=1}^{n}E(\sigma_{ni}^{2})$. Following Hyede and Brown (1970), we verify the following sufficient conditions hold:
\begin{eqnarray}
\sum_{i=1}^{n}E(R_{ni}^{4})/s_{n}^{4}&\rightarrow & 0, \label{conda} \\
E\left(\sum_{i=1}^{n}\sigma_{ni}^{2}-s_{n}^{2}\right)^{2}\bigg/s_{n}^{4} &\rightarrow & 0. \label{condb}
\end{eqnarray}
Not first that
$$
\sigma_{ni}^{2}=\left\{(WZ\alpha)_{i} +2\sum_{j=1}^{i-1}W_{ij}\epsilon_{j}\right\}^{2}\sigma_{\epsilon}^{2}+2W_{ii}\left\{(WZ\alpha)_{i} +2\sum_{j=1}^{i-1}W_{ij}\epsilon_{j}\right\}E\epsilon^{3}+W_{ii}^{2}E(\epsilon^{2}-\sigma_{\epsilon}^{2})^{2}.
$$
Let $E_{z}$ denote $E(\cdot\mid Z)$. It follows that
\begin{eqnarray*}
E_{z}(\sigma_{ni}^{2})&=&\left\{\{(WZ\alpha)_{i}\}^{2} +4\sum_{j=1}^{i-1}W_{ij}^{2}\sigma_{\epsilon}^{2}\right\}\sigma_{\epsilon}^{2}+2W_{ii}(WZ\alpha)_{i} E\epsilon^{3}+W_{ii}^{2}E(\epsilon^{2}-\sigma_{\epsilon}^{2})^{2}.
\end{eqnarray*}
Let $s_{n*}^{2}=\sum_{i=1}^{n}E_{z}(\sigma_{ni}^{2})$. It follows that 
\begin{eqnarray*}
s_{n*}^{2}=\alpha^{t}Z^{t}W^{t}WZ\alpha \sigma_{\epsilon}^{2}+2tr(W^{t}W)\sigma_{\epsilon}^{4}+2\sum_{i=1}^{n}W_{ii}(WZ\alpha)_{i} E\epsilon^{3}+\sum_{i=1}^{n}W_{ii}^{2}\{E(\epsilon^{2}-\sigma_{\epsilon}^{2})^{2}-2\sigma_{\epsilon}^{4}\}.
\end{eqnarray*}
Note further that
$$
E\left(\sum_{i=1}^{n}\sigma_{ni}^{2}-s_{n}^{2}\right)^{2}=E\left(\sum_{i=1}^{n}\sigma_{ni}^{2}-s_{n*}^{2}\right)^{2}+E\left(s_{n*}^{2}-s_{n}^{2}\right)^{2}
$$
We show in the following that $s_{n}\geq c n$ for some constant $c >0$, and
\begin{equation}
\sum_{i=1}^{n}E(R_{ni}^{4})=o(n^{2})
 \mbox{, } E\left(\sum_{i=1}^{n}\sigma_{ni}^{2}-s_{n*}^{2}\right)^{2}=o(n^{2}),
 \mbox{ and } E\left(s_{n*}^{2}-s_{n}^{2}\right)^{2}=o(n^{2}).
\label{conds}
\end{equation}

Note first that
$$
s_{n}^{2}=2E\left\{tr(W^{t}W)-\sum_{i=1}^{n}W_{ii}^{2}\right\}\sigma_{\epsilon}^{4}
+\sum_{i=1}^{n}E\left\{W_{ii}(\epsilon_{i}^{2}-\sigma_{\epsilon}^{2})+(WZ\alpha)_{i}\epsilon_{i}\right\}^{2}.
$$
It follows from the Marchenko-Pastur law that
$$
\frac{1}{n}tr(W^{t}W)\rightarrow \int_{a}^{b}\frac{(x-1)^{2}}{(1+\lambda x)^{4}}d\nu(x),
$$
and from Lemma S.1 at the end of this supplemental material that
$$
\frac{1}{n}\sum_{i=1}^{n}W_{ii}^{2}\rightarrow \left\{\int_{a}^{b} \frac{(x-1)}{(1+\lambda x)^{2}}d\nu(x)\right\}^{2}.
$$
 It follows  from the limit for the maximum eigenvalue of a random matrix (Bai and Silverstein, 1998) that, for sufficiently large $n$, 
$$
s_{n}^{2}\geq  n\left[\int_{a}^{b}\frac{(x-1)^{2}}{(1+\lambda x)^{4}}d\nu(x)-\left\{\int_{a}^{b} \frac{(x-1)}{(1+\lambda x)^{2}}d\nu(x)\right\}^{2}\right]\sigma_{\epsilon}^{4}.
$$

To verify the conditions in (\ref{conds}), note first that
\begin{eqnarray*}
E_{z}(R_{ni}^{4})&=&E_{z}\left[\left\{(WZ\alpha)_{i} +2\sum_{j=1}^{i-1}W_{ij}\epsilon_{j}\right\}\epsilon_{i} +W_{ii}(\epsilon_{i}^{2}-\sigma_{\epsilon}^{2})\right]^{4} \\
&\leq&8\left(E_{z}\left[\left\{(WZ\alpha)_{i} +2\sum_{j=1}^{i-1}W_{ij}\epsilon_{j}\right\}^{4}\right]E\epsilon_{i}^{4} 
+W_{ii}^{4}E\left\{(\epsilon_{i}^{2}-\sigma_{\epsilon}^{2})^{4}\right\}\right). 
\end{eqnarray*}
From the boundedness of these quantities as evident in the verification process, it is sufficient to show that
$$
\sum_{i}E_{z}\left\{(WZ\alpha)_{i} +2\sum_{j=1}^{i-1}W_{ij}\epsilon_{j}\right\}^{4}=o_{p}(n^{2}), 
\mbox{ and } \sum_{i}W_{ii}^{4}=o_{p}(n^{2}),
$$
which in turn are implied by 
\begin{equation}
\sum_{i}\left\{(WZ\alpha)_{i}\right\}^{4}=o_{p}(n^{2}) \mbox{, } \sum_{i}E_{z}\left\{\sum_{j=1}^{i-1}W_{ij}\epsilon_{j}\right\}^{4}=o_{p}(n^{2}), 
\mbox{ and } \sum_{i}W_{ii}^{4}=o_{p}(n^{2}),
\label{cond1}
\end{equation}
The second condition is also equivalent to
$$
\sum_{i=1}^{n}\sum_{j=1}^{i-1}\sum_{l\neq j}^{i-1}W_{ij}^{2}W_{il}^{2}=o_{p}(n^{2}) \mbox{ and } \sum_{i=1}^{n}\sum_{j=1}^{i-1}W_{ij}^{4}=o_{p}(n^{2}).
$$
Note next that
\begin{eqnarray*}
\sum_{i=1}^{n}\sigma_{ni}^{2}-s_{n*}^{2}&=&\sum_{i=1}^{n}\left\{E(\sigma_{ni}^{2}\mid {\cal F}_{ni})-E_{z}(\sigma_{ni}^{2})\right\} \\
&=&4\sigma_{\epsilon}^{2}\sum_{i=1}^{n}(WZ\alpha)_{i}\sum_{j=1}^{i-1}W_{ij}\epsilon_{j}+4E\epsilon^{3}\sum_{i=1}^{n}W_{ii}\sum_{j=1}^{i-1}W_{ij}\epsilon_{j} \\ &&+4\sigma_{\epsilon}^{2}\sum_{i=1}^{n}\left[\left\{\sum_{j=1}^{i-1}W_{ij}\epsilon_{j}\right\}^{2} -\sum_{j=1}^{i-1}W_{ij}^{2}\sigma_{\epsilon}^{2}\right].
\end{eqnarray*}
It follows that
\begin{eqnarray*}
E_{z}\left\{\sum_{i=1}^{n}\sigma_{ni}^{2}-s_{n*}^{2}\right\}^{2} 
&=&E_{z}\bigg[4\sum_{i=1}^{n}\left\{\sigma_{\epsilon}^{2}(WZ\alpha)_{i}+W_{ii}E\epsilon^{3}\right\}\sum_{j=1}^{i-1}W_{ij}\epsilon_{j}  \\ &&+4\sigma_{\epsilon}^{2}\sum_{i=1}^{n}\sum_{j=1}^{i-1}W_{ij}^{2}(\epsilon_{j}^{2}-\sigma_{\epsilon}^{2}) +4\sigma_{\epsilon}^{2}\sum_{i=1}^{n}\sum_{j=1}^{i-1}\sum_{l\neq j}^{i-1}W_{ij}W_{il}\epsilon_{j}\epsilon_{l}\bigg]^{2} \\
&\leq& 3\bigg( E_{z}\left[4\sum_{i=1}^{n}\left\{\sigma_{\epsilon}^{2}(WZ\alpha)_{i}+W_{ii}E\epsilon^{3}\right\}\sum_{j=1}^{i-1}W_{ij}\epsilon_{j}\right]^{2}  \\ &&\hspace{-2cm}+E\left[4\sigma_{\epsilon}^{2}\sum_{i=1}^{n}\sum_{j=1}^{i-1}W_{ij}^{2}(\epsilon_{j}^{2}-\sigma_{\epsilon}^{2})\right]^{2} +E\left[4\sigma_{\epsilon}^{2}\sum_{i=1}^{n}\sum_{j=1}^{i-1}\sum_{l\neq j}^{i-1}W_{ij}W_{il}\epsilon_{j}\epsilon_{l}\right]^{2} \bigg).
\end{eqnarray*}
The first term in the last inequality is $o_{p}(n^{2})$ if conditions (\ref{cond1}) are satisfied. The second term is $o_{p}(n^{2})$ if
\begin{equation}
\sum_{i=1}^{n}\sum_{k=1}^{n}\sum_{j=1}^{i\wedge k -1}W_{ij}^{2}W_{kj}^{2}=o_{p}(n^{2}).
\label{cond2}
\end{equation}
The third term is $o_{p}(n^{2})$ if
\begin{equation}
\sum_{i=1}^{n}\sum_{k=1}^{n}\sum_{j=1}^{i\wedge k -1}\sum_{l\neq j}^{i\wedge k -1}W_{ij}W_{il}W_{kj}W_{kl}=o_{p}(n^{2}).
\label{cond3}
\end{equation}
Conditions (\ref{cond1}), (\ref{cond2}), and (\ref{cond3}) are equivalent to the following conditions
\begin{eqnarray}
\sum_{i=1}^{n}\{(WZ\alpha)_{i}\}^{4}&=&o_{p}(n^{2}), \label{condA}  \\
\sum_{i=1}^{n}\sum_{j=1}^{n}\sum_{l=1}^{n}W_{ij}^{2}W_{il}^{2}&=&o_{p}(n^{2}), \label{condB} \\
\sum_{i=1}^{n}\sum_{k=1}^{n}\sum_{j=1}^{i\wedge k-1}\sum_{l=1}^{i\wedge k-1}W_{ij}W_{il}W_{kj}W_{kl}&=&o_{p}(n^{2}), \label{condC} \\
\sum_{i=1}^{n}\sum_{k=1}^{n}\sum_{j=1}^{i\wedge k-1}W_{ij}^{2}W_{kj}^{2}&=&o_{p}(n^{2}), \label{condD}
\end{eqnarray}
where note that $\sum_{i=1}^{n}W_{ii}^{4}\leq \sum_{i=1}^{n}\sum_{j=1}^{n}\sum_{l=1}^{n}W_{ij}^{2}W_{il}^{2}$. Since $W_{ij}=W_{ji}$,
$$
\sum_{i=1}^{n}\sum_{k=1}^{n}\sum_{j=1}^{i\wedge k-1}W_{ij}^{2}W_{kj}^{2}\leq \sum_{i=1}^{n}\sum_{k=1}^{n}\sum_{j=1}^{n}W_{ij}^{2}W_{kj}^{2}
=\sum_{j=1}^{n}\sum_{i=1}^{n}\sum_{k=1}^{n}W_{ji}^{2}W_{jk}^{2}.
$$
(\ref{condB}) implies (\ref{condD}).  Note that $W_{ij}=\sum_{s=1}^{n}\delta_{s}U_{is}U_{js}$, where $U=(U_{ik})$ is an orthonormal matrix and $\delta_{s}, s=1,\cdots,n$ are eigenvalues of $W$. Plugging this in (\ref{condB}), it follows that
\begin{eqnarray*}
\sum_{i=1}^{n}\sum_{j=1}^{n}\sum_{l=1}^{n}W_{ij}^{2}W_{il}^{2}&=&
\sum_{i=1}^{n}\sum_{j=1}^{n}\sum_{l=1}^{n}\sum_{s=1}^{n}\delta_{s}U_{is}U_{js}
\sum_{t=1}^{n}\delta_{t}U_{it}U_{jt}\sum_{u=1}^{n}\delta_{u}U_{iu}U_{ku}\sum_{v=1}^{n}\delta_{v}U_{iv}U_{kv} \\
&=&\sum_{s=1}^{n}\sum_{t=1}^{n}\sum_{u=1}^{n}\sum_{v=1}^{n}\delta_{s}\delta_{t}\delta_{u}\delta_{v}\sum_{i=1}^{n}
U_{is}U_{it}U_{iu}U_{iv}\sum_{j=1}^{n}U_{js}U_{jt}\sum_{k=1}^{n}U_{ku}U_{kv} \\
&=&\sum_{s=1}^{n}\sum_{u=1}^{n}\delta_{s}^{2}\delta_{u}^{2}\sum_{i=1}^{n}U_{is}^{2}U_{iu}^{2} =\sum_{i=1}^{n}\left(\sum_{j=1}^{n}\delta_{j}^{2}U_{ij}^{2}\right)^{2}=\sum_{i=1}^{n}\left\{(W^{2})_{ii}\right\}^{2} \\
&\leq&\sum_{i=1}^{n}\sum_{j=1}^{n}\left\{(W^{2})_{ij}\right\}^{2}=tr(W^{4})=O_{p}(n)
\end{eqnarray*}
Note also that
\begin{eqnarray*}
\sum_{i=1}^{n}\sum_{k=1}^{n}\sum_{j=1}^{i\wedge k-1}\sum_{l=1}^{i\wedge k-1}W_{ij}W_{il}W_{kj}W_{kl}
&=&\sum_{i=1}^{n}\sum_{k=1}^{i-1}\sum_{j=1}^{k-1}\sum_{l=1}^{k-1}W_{ij}W_{il}W_{kj}W_{kl} \\
&&\hspace{-1.8in}+\sum_{i=1}^{n}\sum_{k=i+1}^{n}\sum_{j=1}^{i-1}\sum_{l=1}^{i-1}W_{ij}W_{il}W_{kj}W_{kl}
+\sum_{i=1}^{n}\sum_{j=1}^{i-1}\sum_{l=1}^{i-1}W_{ij}^{2}W_{il}^{2} \\
&&\hspace{-2in}=2\sum_{i=1}^{n}\sum_{k=1}^{i-1}\sum_{j=1}^{k-1}\sum_{l=1}^{k-1}W_{ij}W_{il}W_{kj}W_{kl}+O_{p}(n) \\
&&\hspace{-2in}=4\sum_{i=1}^{n}\sum_{k=1}^{i-1}\sum_{j=1}^{k-1}\sum_{l=1}^{j-1}W_{ij}W_{il}W_{kj}W_{kl}
+O_{p}(n).
\end{eqnarray*}
Note also that
\begin{eqnarray*}
tr(W^{4})&=&2\sum_{i=1}^{n}\sum_{k=1}^{n}\sum_{j=1}^{n}\sum_{l=1}^{j-1}W_{ij}W_{jk}W_{kl}W_{li}
+\sum_{i=1}^{n}\sum_{k=1}^{n}\sum_{j=1}^{n}W_{ij}^{2}W_{jk}^{2} \\
&=&4\sum_{i=1}^{n}\sum_{k=1}^{i-1}\sum_{j=1}^{n}\sum_{l=1}^{j-1}W_{ij}W_{jk}W_{kl}W_{li}+O(n) \\
&=&8\sum_{i=1}^{n}\sum_{j=1}^{i-1}\sum_{k=1}^{j-1}\sum_{l=1}^{j-1}W_{ij}W_{jk}W_{kl}W_{li} 
+4\sum_{i=1}^{n}\sum_{k=1}^{i-1}\sum_{l=1}^{i-1}W_{ii}W_{ik}W_{kl}W_{li} +O(n)   \\
&=&16\sum_{i=1}^{n}\sum_{j=1}^{i-1}\sum_{k=1}^{j-1}\sum_{l=1}^{k-1}W_{ij}W_{jk}W_{kl}W_{li} 
+8\sum_{i=1}^{n}\sum_{j=1}^{i-1}\sum_{k=1}^{j-1}W_{ij}W_{jk}W_{kk}W_{ki}  \\
&&+8\sum_{i=1}^{n}\sum_{k=1}^{i-1}\sum_{l=1}^{k-1}W_{ii}W_{ik}W_{kl}W_{li}
+4\sum_{i=1}^{n}\sum_{k=1}^{i-1}W_{ii}W_{ik}^{2}W_{kk}+O_{p}(n)   .
\end{eqnarray*}
Since
$$
\sum_{i=1}^{n}\sum_{k=1}^{i-1}W_{ii}W_{ik}^{2}W_{kk}=\frac{1}{2}\left(\sum_{i=1}^{n}\sum_{k=1}^{n}W_{ii}W_{ik}^{2}W_{kk}-\sum_{i=1}^{n}W_{ii}^{4}\right),
$$
and 
$$
\sum_{i=1}^{n}\sum_{k=1}^{n}W_{ii}W_{ik}^{2}W_{kk}\leq \left(\sum_{i=1}^{n}\sum_{k=1}^{n}W_{ii}^{2}W_{kk}^{2}\sum_{i=1}^{n}\sum_{k=1}^{n}W_{ik}^{4}\right)^{1/2}=O_{p}(n^{3/2}),
$$
it follows that
$$
\sum_{i=1}^{n}\sum_{k=1}^{i-1}W_{ii}W_{ik}^{2}W_{kk}=O_{p}(n^{3/2}).
$$
Note further that
\begin{eqnarray*}
\sum_{i=1}^{n}\sum_{j=1}^{i-1}\sum_{k=1}^{j-1}W_{ij}W_{jk}W_{kk}W_{ki} 
&=&\frac{1}{8}\sum_{i=1}^{n}\sum_{j=1}^{n}\sum_{k=1}^{n}W_{ij}W_{jk}W_{kk}W_{ki}-\frac{1}{8}\sum_{i=1}^{n}\sum_{j=1}^{n}W_{ij}^{2}W_{jj}^{2} \\
&&-\frac{3}{4}\sum_{i=1}^{n}\sum_{j=1}^{i-1}W_{ii}W_{ij}^{2}W_{jj} 
\end{eqnarray*} 
Since
$$
\sum_{i=1}^{n}\sum_{j=1}^{n}\sum_{k=1}^{n}W_{ij}W_{jk}W_{kk}W_{ki}\leq \left(\sum_{i=1}^{n}\sum_{j=1}^{n}\sum_{k=1}^{n}W_{ij}^{2}W_{kk}^{2}
\sum_{i=1}^{n}\sum_{j=1}^{n}\sum_{k=1}^{n}W_{jk}^{2}W_{ki}^{2}\right)^{1/2}=O_{p}(n^{3/2}),
$$
it follows that
$$
\sum_{i=1}^{n}\sum_{j=1}^{i-1}\sum_{k=1}^{j-1}W_{ij}W_{jk}W_{kk}W_{ki} =O_{p}(n^{3/2}).
$$
Put together, we have
$$
\sum_{i=1}^{n}\sum_{j=1}^{i-1}\sum_{k=1}^{j-1}\sum_{l=1}^{k-1}W_{ij}W_{jk}W_{kl}W_{li} =O_{p}(n^{3/2}).
$$
This suggests condition (\ref{condB}) holds.

Finally, $\sum_{i=1}^{n}\left\{(WZ\alpha)_{i}\right\}^{4}=\sigma_{s}^{4}\sum_{i=1}^{n}\left\{(WZu)_{i}\right\}^{4}$, where $||u||_{2}=1$. Furthermore,
$$
\sum_{i=1}^{n}\left\{(WZu)_{i}\right\}^{4}\leq \max_{1\leq i\leq n}(WZu)_{i}^{2}\sum_{i=1}^{n}(WZu)_{i}^{2}.
$$
Since, from Corollary 2 of Bai et al (2007),
$$
\frac{1}{p}\sum_{i=1}^{n}(WZu)_{i}^{2}\rightarrow \xi^{-1}\int \frac{(x-1)^{2}x}{(1+\lambda x)^{4}}d\nu(x),
$$
it is sufficient to show that
$$
\max_{1\leq i\leq n}(WZu/\sqrt{p})_{i}^{2}=o_{p}(1).
$$
Note that, for $\lambda>0$, 
$$
\max_{1\leq i\leq n}(WZu/\sqrt{p})_{i}^{2}\leq ||W^{2}||\max_{1\leq i\leq n}(Zu/\sqrt{p})_{i}^{2}=\frac{1}{16\lambda^{2}(\lambda+1)^{2}}\max_{1\leq i\leq n}(Zu/\sqrt{p})_{i}^{2}.
$$
Furthermore,
$$
 E\max_{1\leq i\leq n}(Zu/\sqrt{p})_{i}^{2}\leq E\left(\frac{1}{p^{2}}\sum_{i=1}^{n}(Zu)_{i}^{4}\right)^{1/2}\leq \left[E\left\{p^{-2}\sum_{i=1}^{n}(Zu)_{i}^{4}\right\}\right]^{1/2}.
$$
Direct computation can show easily that
$$
E\left\{p^{-2}\sum_{i=1}^{n}(Zu)_{i}^{4}\right\}=O(p^{-2}).
$$

Lastly, to show $E(s_{n*}^{2}-s_{n}^{2})^{2}=o(n^{2})$, it is sufficient to show that
\begin{eqnarray*}
E\left\{\alpha^{t}Z^{t}W^{t}WZ\alpha-E(\alpha^{t}Z^{t}W^{t}WZ\alpha)\right\}^{2}&=&o(n^{2}), \\
E\left[tr(W^{t}W)-E\{tr(W^{t}W)\}\right]&=&o(n^{2}), \\
E\left\{\sum_{i=1}^{n}(W_{ii}^{2}-EW_{ii}^{2})\right\}^{2}&=&o(n^{2}), \\
E\left(\sum_{i=1}^{n}\left[W_{ii}(WZ\alpha)_{i}-E\{W_{ii}(WZ\alpha)_{i}\}\right]\right)^{2}&=&o(n^{2}).
\end{eqnarray*}
The first one is true due to Theorem 1 of Bai et al (2007). The second is true due to the Marchenko-Pastur Law. The third one is true due to Lemma S.1. For the last one, note that
$$ 
\sum_{i=1}^{n}W_{ii}(WZ\alpha)_{i}=\sum_{i=1}^{n}\left(W_{ii}-\int g(x)d\nu(x)\right)(WZ\alpha)_{i}+\int g(x)d\nu(x)\sum_{i=1}^{n}(WZ\alpha)_{i}.
$$
It follows that
$$
\left|\sum_{i=1}^{n}W_{ii}(WZ\alpha)_{i}-\int g(x)d\nu(x)\sum_{i=1}^{n}(WZ\alpha)_{i}\right|=\max_{1\leq i\leq n}\left|W_{ii}-\int g(x)d\nu(x)\right|\sum_{i=1}^{n}|(WZ\alpha)_{i}|
$$
Furthermore,
$$
\frac{1}{n}\sum_{i=1}^{n}|(WZ\alpha)_{i}|\leq \left\{\frac{1}{n}\sum_{i=1}^{n}(WZ\alpha)_{i}^{2}\right\}^{1/2}\rightarrow \xi^{-1}\int \frac{(x-1)^{2}x}{(1+\lambda x)^{4}}d\nu(x).
$$
The last one follows. 

\section{A Lemma for the proofs}

\noindent {\em Lemma S.1.} For any fixed $k$, 
$$
\frac{1}{n}\sum_{i=1}^{n}W_{ii}^{k}\rightarrow \left\{\int g(x) d\nu(x)\right\}^{k}, a.s.
$$
where $g(\lambda)$ is an eigenvalue of $W$ if $\lambda$ is an eigenvalue of $M$.

\vspace{0.3cm}

\noindent Proof: For $k=1$,
$$
\frac{1}{n}\sum_{i=1}^{n}W_{ii}=\frac{1}{n}tr(W)=\frac{1}{n}\sum_{i=1}^{n}g(\lambda_{i})\rightarrow \int g(x)d\nu(x), a.s.
$$
by the Marchenko-Pastur law or Corollary 1 of Theorem 1 in Bai et al (2007). Suppose that the result holds for $k=m-1$.  For $k=m\geq 2$,
$$
\left|\frac{1}{n}\sum_{i=1}^{n}W_{ii}^{m}-\left\{ \int g(x)d\nu(x)\right\}^{m}\right|
\leq \max_{1\leq i\leq n}\left|W_{ii}- \int g(x)d\nu(x)\right| R_{m},
$$
where 
$$
R_{m}=\sum_{j=0}^{m-1}\frac{1}{n}\sum_{i=1}^{n}|W_{ii}|^{m-1-j}\left|\int g(x)d\nu(x)\right|^{j}.
$$
From the argument of Remark 2 of Theorem 1 in Bai et al (2007), it follows that
$$
\max_{1\leq i\leq n}\left| W_{ii}- \int g(x)d\nu(x)\right|\rightarrow 0,
$$
as $n\rightarrow \infty$. For any $j\geq 1$
$$
\frac{1}{n}\sum_{i=1}^{n}|W_{ii}|^{j}\leq C_{j1}\max_{1\leq i\leq n}\left|W_{ii}-\int g(x) d\nu(x)\right|+C_{j2},
$$
when $n$ is sufficient large, where $C_{j1}, C_{j2}$ are constants depend on $\int g(x)d\mu(x)$, but independent of $n$. The Lemma follows.

\section{Tabulated Simulation Results} 

\begin{table}
\caption{Simulation results on the explained variation with normal covariates and normal random error, and a sample size of $200$.}
%\begin{center}
\begin{tabular}{ccl lll cc} \hline \hline
         &               &                      &             &  var              &  e.var                     &  (99, 95, 90)\% & (99, 95, 90)\%  \\
    p    &  $r^{2}$ &   method       &  est.     & $\times 10^{3}$  &  $\times 10^{3}$      &   CI coverage     & CI length     \\   \hline
 100  &    0.2     &   EigenPrism     & 0.201  & 8.79   &     ---    &   (99.9, 98.2, 95.2) &  (0.426, 0.364, 0.322)   \\ 
         &              &   Est.Equ         & 0.201  & 7.05   & 6.88     &   (99.4, 95.1, 89.2) &  (0.386, 0.306, 0.261)   \\ 
         &              &   Est.Equ(N)    & 0.201  & 7.05   & 6.88     &   (99.3, 94.8, 89.4) &  (0.386, 0.306, 0.261)   \\ 
         &              &   GCTA            & 0.201  & 8.19   & 6.97     &   (98.6, 94.5, 91.1) &  (0.390, 0.317, 0.273)    \\ 

         &    0.5     &   EigenPrism    & 0.496  & 4.56   &     ---    &   (99.5, 96.0, 91.6) &  (0.391, 0.292, 0.243)   \\ 
         &              &   Est.Equ         & 0.495  & 4.43   & 4.98     &   (99.0, 95.4, 91.5) &  (0.360, 0.274, 0.230)   \\ 
         &              &   Est.Equ(N)    & 0.495  & 4.43   & 4.30     &   (98.0, 93.8, 89.8) &  (0.336, 0.255, 0.214)   \\ 
         &              &   GCTA            & 0.494  & 4.81   & 6.56     &   (99.6, 97.3, 94.4) &  (0.423, 0.317, 0.265)    \\ 

         &    0.8     &   EigenPrism     & 0.799  & 1.09   &     ---    &   (97.6, 91.5, 85.8) &  (0.156, 0.117, 0.097)   \\ 
         &              &   Est.Equ         & 0.798  & 1.14   & 3.58     &   (99.7, 99.0, 98.5) &  (0.297, 0.229, 0.192)   \\ 
         &              &   Est.Equ(N)    & 0.798  & 1.14   & 0.81     &   (97.2, 90.1, 82.5) &  (0.145, 0.110, 0.092)   \\ 
         &              &   GCTA            & 0.797  & 1.79   & 2.31     &   (99.4, 96.8, 92.9) &  (0.255, 0.187, 0.156)    \\ 

  200 &    0.2      &   EigenPrism   & 0.216  & 15.8   &     ---    &   (99.6, 97.3, 94.3) &  (0.594, 0.485, 0.422)   \\ 
         &              &   Est.Equ         & 0.201  & 11.5   & 11.9     &   (99.6, 97.5, 90.3) &  (0.467, 0.377, 0.324)   \\ 
         &              &   Est.Equ(N)    & 0.201  & 11.5   & 11.9     &   (99.6, 97.5, 90.3) &  (0.467, 0.376, 0.324)   \\ 
         &              &   GCTA            & 0.201  &  11.7  & 10.4     &   (99.6, 96.6, 89.7) &  (0.453, 0.376, 0.326)    \\ 

         &    0.5     &   EigenPrism    & 0.494  & 10.6   &     ---    &   (99.7, 99.3, 98.1) &  (0.759, 0.586, 0.492)   \\ 
         &              &   Est.Equ         & 0.491  & 9.92   & 9.05     &   (98.6, 94.3, 88.8) &  (0.484, 0.369, 0.310)   \\ 
         &              &   Est.Equ(N)    & 0.491  & 9.92   & 8.21     &   (97.7, 93.2, 86.5) &  (0.461, 0.352, 0.295)   \\ 
         &              &   GCTA            & 0.490  & 10.6   & 10.4     &   (98.1, 94.9, 90.3) &  (0.528, 0.398, 0.334)    \\ 
 
         &    0.8     &   EigenPrism    & 0.799  & 3.56   &     ---    &   (  100,  100, 99.9) &  (0.587, 0.492, 0.439)   \\ 
         &              &   Est.Equ         & 0.799  & 2.92   & 5.36     &   (99.7, 98.8, 96.7) &  (0.355, 0.279, 0.236)   \\ 
         &              &   Est.Equ(N)    & 0.799  & 2.92   & 2.10     &   (95.2, 89.2, 83.3) &  (0.231, 0.176, 0.148)   \\ 
         &              &   GCTA            & 0.796  & 4.80   & 5.10     &   (99.2, 95.8, 90.3) &  (0.375, 0.278, 0.232)    \\ 

  800  &    0.2    &   EigenPrism    & 0.236  & 32.1   &     ---    &   (99.8, 96.8, 92.1) &  (0.772, 0.636, 0.554)   \\ 
         &              &   Est.Equ         & 0.207  & 31.0   & 41.3     &   (99.6, 97.8, 95.7) &  (0.719, 0.586, 0.510)   \\ 
         &              &   Est.Equ(N)    & 0.207  & 31.0   & 41.3     &   (99.5, 97.8, 95.7) &  (0.719, 0.587, 0.510)   \\ 
         &              &   GCTA            & 0.206  & 30.7   & 27.8     &   (99.8, 98.1, 96.4) &  (0.685, 0.576, 0.508)    \\ 

         &    0.5     &   EigenPrism    & 0.519  & 34.5   &     ---    &   (99.7, 98.5, 95.8) &  (0.909, 0.774, 0.676)   \\ 
         &              &   Est.Equ         & 0.501  & 37.5   & 36.2     &   (99.3, 95.4, 89.3) & (0.830, 0.680, 0.586)   \\ 
         &              &   Est.Equ(N)    & 0.501  & 37.5   & 35.1     &   (99.1, 93.7, 88.3) & (0.822, 0.671, 0.578)   \\ 
         &              &   GCTA            & 0.498  & 38.8   & 34.7     &   (99.7, 95.8, 89.5) & (0.851, 0.705, 0.609)    \\ 

         &    0.8     &   EigenPrism     & 0.787  & 23.9   &     ---    &   ( 100, 99.1, 97.1) &  (0.767, 0.631, 0.553)   \\ 
         &              &   Est.Equ         & 0.784  & 24.7   &  27.1    &   (  100,  100, 96.5) &  (0.617, 0.520, 0.456)   \\
         &              &   Est.Equ(N)    & 0.784  & 24.7   &  22.4    &   (  100, 89.7, 84.8) &  (0.573, 0.485, 0.423)   \\ 
         &              &   GCTA            & 0.777  & 27.8   & 25.9     &    (98.8, 97.1, 94.5) &  (0.745, 0.584, 0.500)    \\ 
\hline\hline
\end{tabular}
%\end{center}
\label{Sim2hnorm}
\end{table}
 
 \begin{table}
\caption{Simulation results on the explained variation with squared normal covariates and cubed normal random error, and a sample size of $200$.}
%\begin{center}
\begin{tabular}{ccl lll cc} \hline \hline
         &               &                      &             &  var              &  e.var                     &  (99, 95, 90)\% & (99, 95, 90)\%  \\
    p    &  $r^{2}$ &   method       &  est.     & $\times 10^{3}$  &  $\times 10^{3}$      &   CI coverage     & CI length     \\   \hline
  100  &    0.2     &   EigenPrism    & 0.221  & 14.6   &     ---    &   (98.5, 93.3, 86.7) &  (0.430, 0.358, 0.312)   \\ 
         &              &   Est.Equ         & 0.220  & 12.3   & 21.9     &   (99.4, 97.6, 94.2) &  (0.575, 0.468, 0.405)   \\ 
         &              &   Est.Equ(N)    & 0.220  & 12.3   & 6.88     &   (96.4, 85.4, 75.6) &  (0.384, 0.302, 0.257)   \\ 
         &              &   GCTA            & 0.219  & 12.2   & 6.84     &   (95.6, 87.1, 79.3) &  (0.391, 0.313, 0.268)    \\ 

         &    0.5     &   EigenPrism     & 0.523  & 15.2   &     ---    &   (84.6, 71.8, 63.4) &  (0.363, 0.273, 0.228)   \\ 
         &              &   Est.Equ         & 0.522  & 15.0   & 28.4     &   (96.4, 92.5, 89.7) &  (0.708, 0.570, 0.488)   \\ 
         &              &   Est.Equ(N)    & 0.522  & 15.0   & 4.08     &   (78.1, 67.3, 59.0) &  (0.320, 0.244, 0.205)   \\ 
         &              &   GCTA            & 0.520  & 15.7   & 6.07     &   (89.6, 78.7, 70.5) &  (0.404, 0.303, 0.254)    \\ 

         &    0.8     &   EigenPrism     & 0.806  & 5.42   &     ---    &   (68.7, 52.3, 44.9) &  (0.151, 0.112, 0.093)   \\ 
         &              &   Est.Equ         & 0.806  & 5.47   & 9.91     &   (99.8, 98.1, 95.1) &  (0.417, 0.341, 0.294)   \\ 
         &              &   Est.Equ(N)    & 0.806  & 5.47   & 0.84     &   (61.9, 49.3, 42.4) &  (0.140, 0.107, 0.089)   \\ 
         &              &   GCTA            & 0.803  & 6.15   & 2.25     &   (88.4, 77.6, 68.9) &  (0.246, 0.181, 0.151)    \\ 

  200 &    0.2     &   EigenPrism     & 0.235  & 21.7   &     ---    &   (98.4, 93.3, 87.3) &  (0.602, 0.486, 0.419)   \\ 
         &              &   Est.Equ         & 0.223  & 18.8   & 28.3     &   (99.1, 96.4, 93.9) &  (0.632, 0.514, 0.446)   \\ 
         &              &   Est.Equ(N)    & 0.223  & 18.8   & 11.9     &   (96.8, 90.5, 78.5) &  (0.471, 0.375, 0.322)   \\ 
         &              &   GCTA            & 0.222  & 18.6   & 10.2     &   (97.9, 89.8, 78.7) &  (0.458, 0.373, 0.322)    \\ 

         &    0.5     &   EigenPrism     & 0.526  & 20.4   &     ---    &   (99.2, 96.2, 91.7) &  (0.737, 0.578, 0.489)   \\ 
         &              &   Est.Equ         & 0.521  & 19.7   & 34.9     &   (98.0, 93.9, 90.7) &  (0.767, 0.626, 0.539)   \\ 
         &              &   Est.Equ(N)    & 0.521  & 19.7   & 7.81     &   (85.8, 75.6, 67.6) &  (0.440, 0.337, 0.284)   \\ 
         &              &   GCTA            & 0.519  & 20.1   & 9.72     &   (93.3, 82.3, 75.0) &  (0.507, 0.383, 0.322)    \\ 
 
         &    0.8     &   EigenPrism  & 0.800  & 8.32   &     ---    &   (99.9, 99.3, 98.3) &  (0.583, 0.484, 0.428)   \\ 
         &              &   Est.Equ         & 0.803  & 8.05   & 13.2     &   (99.8, 97.8, 93.5) &  (0.453, 0.374, 0.326)   \\ 
         &              &   Est.Equ(N)    & 0.803  & 8.05   & 2.21     &   (80.2, 69.5, 61.1) &  (0.228, 0.174, 0.146)   \\ 
         &              &   GCTA            & 0.798  & 9.88   & 5.05     &   (93.6, 85.1, 77.5) &  (0.368, 0.274, 0.228)    \\ 

  800  &    0.2     &   EigenPrism     & 0.284  & 40.9   &     ---    &   (99.2, 96.2, 91.7) &  (0.737, 0.578, 0.489)   \\ 
         &              &   Est.Equ         & 0.253  & 40.2   & 68.0     &   (99.6, 97.5, 95.5) &  (0.861, 0.728, 0.641)   \\ 
         &              &   Est.Equ(N)    & 0.253  & 40.2   & 41.1     &   (98.2, 95.0, 91.9) &  (0.750, 0.613, 0.532)   \\ 
         &              &   GCTA            & 0.253  & 40.3   & 29.2     &   (98.7, 95.8, 93.2) &  (0.718, 0.603, 0.529)    \\ 

         &    0.5     &   EigenPrism     & 0.529  & 45.7   &     ---    &   (99.5, 95.5, 89.8) &  (0.886, 0.751, 0.656)   \\ 
         &              &   Est.Equ         & 0.512  & 49.5   & 70.5     &   (99.3, 96.8, 92.8) &  (0.915, 0.802, 0.715)   \\ 
         &              &   Est.Equ(N)    & 0.512  & 49.5   & 35.0     &   (95.6, 88.8, 81.3) &  (0.797, 0.650, 0.560)   \\ 
         &              &   GCTA            & 0.508  & 50.6   & 33.2     &   (97.6, 89.5, 83.7) &  (0.827, 0.684, 0.591)    \\ 

         &    0.8     &   EigenPrism     & 0.791  & 27.9   &     ---    &   (99.9, 97.8, 95.5) &  (0.756, 0.620, 0.542)   \\ 
         &              &   Est.Equ         & 0.790  & 28.8   & 56.7     &   (  100,  100, 98.5) &  (0.767, 0.626, 0.539)   \\ 
         &              &   Est.Equ(N)    & 0.790  & 28.8   & 25.2     &   (  100, 85.7, 79.7) &  (0.558, 0.472, 0.412)   \\ 
         &              &   GCTA            & 0.783  & 32.1   & 25.2     &   (98.3, 95.4, 92.8) &  (0.733, 0.571, 0.489)    \\ 
\hline\hline
\end{tabular}
%\end{center}
\label{Sim2hnon}
\end{table}
 
\begin{table}
\caption{Simulation results on the explained variation with normal covariates and normal random error, and a sample size of $800$.}
%\begin{center}
\begin{tabular}{ccl lll cc} \hline \hline
         &               &                      &             &  var              &  e.var                     &  (99, 95, 90)\% & (99, 95, 90)\%  \\
    p    &  $r^{2}$ &   method       &  est.     & $\times 10^{3}$  &  $\times 10^{3}$  &   CI coverage     & CI length     \\   \hline
 400  &    0.0      &   EigenPrism     & 0.019  & 2.36   &     ---    &  (99.9, 99.7, 98.0)    & (0.160, 0.125, 0.106)   \\ 
         &               &   Est.Equ         & 0.014  & 0.44   & 1.33     &   (100, 98.6, 96.3)    & (0.108, 0.085, 0.074) \\  
         &               &   Est.Equ(N)    & 0.014  & 0.44   & 1.33     &   (100, 98.6, 96.3)    & (0.108, 0.085, 0.074)   \\ 

         &    0.2      &   EigenPrism    & 0.200  & 2.23   &     ---     &  (99.8, 98.1, 94.8)    & (0.286, 0.222, 0.187)   \\ 
         &               &   Est.Equ         & 0.200  & 1.78   & 1.72     &   (98.1, 94.8, 89.7)   & (0.213, 0.162, 0.136) \\  
         &               &   Est.Equ(N)    & 0.200  & 1.78   & 1.70     &   (98.1, 94.7, 89.2)   & (0.212, 0.162, 0.136)   \\ 

         &    0.5      &   EigenPrism    & 0.498  & 1.22   &     ---    &  (99.0, 95.0, 90.6)    &  (0.186, 0.141, 0.118)   \\ 
         &               &   Est.Equ        &  0.498  & 1.21   & 1.50     &   (99.8, 97.2, 93.2)   & (0.199, 0.152, 0.127) \\ 
         &               &   Est.Equ(N)    & 0.498  & 1.21   & 1.05     &   (98.3, 94.3, 87.3)   & (0.167, 0.127, 0.107)   \\ 
 
         &    0.8      &   EigenPrism    & 0.799  & 0.25   &     ---    &   (97.8, 91.6, 86.3)    & (0.074, 0.056, 0.047)  \\ 
         &               &   Est.Equ         & 0.799  & 0.26   & 1.75     &   (100, 100, 100)        & (0.215, 0.164, 0.137) \\ 
         &               &   Est.Equ(N)    & 0.799  & 0.26   & 0.20     &   (97.2, 91.1, 85.9)    & (0.072, 0.055, 0.046)   \\ 

 800  &    0.0      &   EigenPrism    & 0.031  & 1.71   &     ---     &   (99.3, 95.6, 90.7)   &  (0.197, 0.150, 0.126)  \\ 
         &              &   Est.Equ        & 0.021  & 0.96   & 2.62     &   (99.7. 98.4, 96.1)    & (0.153, 0.121, 0.104) \\  
         &              &   Est.Equ(N)    & 0.021  & 0.96   & 2.62     &   (99.7. 98.4, 96.1)    & (0.153, 0.121, 0.104) \\  
 
         &    0.2     &   EigenPrism   & 0.204  & 4.26   &     ---   &   (99.6, 97.6, 94.4)     &  (0.366, 0.288, 0.244)  \\ 
         &              &   Est.Equ        & 0.200  & 2.96   & 2.95     &   (98.6. 94.7, 90.5)    & (0.276, 0.212, 0.178) \\  
         &              &   Est.Equ(N)    & 0.200  & 2.96   & 2.94     &   (98.6, 94.7, 90.6)    & (0.275, 0.211, 0.178)   \\ 
  
         &    0.5      &   EigenPrism    & 0.501  & 2.33   &     ---     &   (100, 99.8, 99.0)     & (0.387, 0.295, 0.247)   \\ 
         &               &   Est.Equ        & 0.500   & 2.03   & 2.44     &   (99.5, 96.4, 92.7)   & (0.254, 0.193, 0.162) \\ 
         &              &   Est.Equ(N)    & 0.500   & 2.03   & 1.97     &   (98.7, 94.0, 89.8)   & (0.228, 0.174, 0.146)   \\ 

         &    0.8     &   EigenPrism    & 0.799  & 0.91   &     ---     &  (100, 100, 100)        & (0.378, 0.294, 0.247)   \\ 
         &              &   Est.Equ         & 0.799  & 0.70   & 2.23     &   (100, 99.8, 99.7)    & (0.243, 0.185, 0.155) \\ 
         &              &   Est.Equ(N)    & 0.799  & 0.70   & 0.51     &   (96.8, 90.0, 83.1)  & (0.116, 0.088, 0.074)   \\ 

3200  &    0.0     &   EigenPrism     & 0.047  & 4.27   &     ---     &  (99.4, 96.3, 91.1)   & (0.296, 0.227, 0.190)   \\ 
         &              &   Est.Equ         & 0.038  & 3.28   & 10.2      &  (99.5, 98.2, 96.3)   & (0.298, 0.235, 0.202)   \\ 
         &              &   Est.Equ(N)    & 0.038  & 3.28   & 10.2      &  (99.5, 98.2, 96.3)   & (0.298, 0.235, 0.202)   \\ 

         &    0.2     &   EigenPrism    & 0.203  & 10.5   &     ---     &  (98.8, 96.4, 91.2)   & (0.479, 0.388, 0.334)   \\ 
         &              &   Est.Equ         & 0.195  & 9.82   & 10.4      &  (99.6, 97.6, 90.7)   & (0.442, 0.357, 0.307)   \\ 
         &              &   Est.Equ(N)    & 0.195  & 9.82   & 10.4      &  (99.6, 97.6, 90.7)   & (0.442, 0.356, 0.307)   \\ 

         &    0.5     &   EigenPrism    & 0.508  & 8.96   &     ---    &   (99.3, 97.5, 94.3)  & (0.576, 0.439, 0.369)   \\ 
         &              &   Est.Equ         & 0.503  & 9.04   & 9.18     &   (98.6, 95.1, 90.8)  & (0.492, 0.375, 0.315) \\ 
         &              &   Est.Equ(N)    & 0.503  & 9.04   & 8.89     &   (98.5, 94.9, 89.8)  & (0.485, 0.369, 0.310)   \\ 

         &    0.8     &   EigenPrism    & 0.801  & 6.36   &     ---    &    (100, 99.3,98.2)    & (0.483, 0.396, 0.343)   \\ 
         &              &   Est.Equ         & 0.801  & 6.14   & 6.83     &   (99.6, 96.1, 91.4)  & (0.391, 0.310, 0.264) \\ 
         &              &   Est.Equ(N)    & 0.801  & 6.14   & 5.47     &   (98.0, 92.6, 87.5)  & (0.360, 0.282, 0.238)   \\ 
\hline\hline
\end{tabular}
%\end{center}
\label{Sim8hnorm}
\end{table}
 
 \begin{table}
\caption{Simulation results on the explained variation with squared normal covariates and cubed normal random error, and a sample size of $800$.}
%\begin{center}
\begin{tabular}{ccl lll cc} \hline \hline
         &               &                      &             &  var              &  e.var                     &  (99, 95, 90)\% & (99, 95, 90)\%  \\
    p    &  $r^{2}$ &   method       &  est.     & $\times 10^{3}$  &  $\times 10^{3}$      &   CI coverage     & CI length     \\   \hline

400  &    0.0      &   EigenPrism    & 0.000  & 2.89   &     ---   &   (99.9, 98.7, 96.9)    & (0.161, 0.125, 0.106)     \\ 
         &              &   Est.Equ         & 0.015  & 0.58   & 1.69   &   (99.7, 98.5, 96.1)    & (0.121, 0.095, 0.082) \\ 
         &              &   Est.Equ(N)    & 0.015  & 0.58   & 1.33   &   (99.4, 97.2, 94.6)    &  (0.109, 0.086, 0.074)   \\ 

         &    0.2     &   EigenPrism    & 0.204  & 3.53   &     ---   &   (98.3, 93.4, 88.7)    & (0.281, 0.219, 0.185)     \\ 
         &               &   Est.Equ        & 0.203  & 3.12   &  6.86   &   (99.6, 98.4, 96.4)   & (0.379, 0.300, 0.256) \\ 
         &              &   Est.Equ(N)    & 0.203  & 3.12   & 1.70   &   (94.4, 86.4, 78.2)    &  (0.211, 0.161, 0.135)   \\ 
 
         &    0.5      &   EigenPrism    & 0.505  & 4.38   &     ---    &   (82.9, 69.7, 61.1)   & (0.183, 0.138, 0.116)   \\ 
         &               &   Est.Equ        & 0.505   & 4.37   & 9.81     &   (99.2, 96.7, 94.6)   & (0.474, 0.364, 0.304) \\ 
         &               &   Est.Equ(N)    & 0.505   & 4.37   & 1.04     &   (77.0, 64.9, 57.3)  & (0.165, 0.125, 0.105)   \\ 
 
         &    0.8      &   EigenPrism    & 0.801  & 1.58   &     ---     &  (65.3, 53.2, 45.9)     & (0.074, 0.056, 0.047) \\ 
         &               &   Est.Equ        & 0.801  &  1.59   & 3.74     &   (100,   99.7, 99.1)    & (0.300, 0.229, 0.193) \\ 
         &               &   Est.Equ(N)    & 0.801  &  1.59   & 0.20     &   (63.6, 52.5, 44.8)    & (0.071, 0.054, 0.046)   \\ 
 
  800  &  0.0      &   EigenPrism    & 0.034  & 2.36   &     ---    &    (98.6, 94.0, 88.9)   &  (0.198, 0.150, 0.125)   \\ 
         &              &   Est.Equ         & 0.022  & 1.13   & 3.33     &   (99.9, 98.5, 96.4)    & (0.170, 0.134, 0.116) \\ 
         &              &   Est.Equ(N)    & 0.022  & 1.13   & 2.94     &   (99.4, 97.4, 94.5)    & (0.154, 0.122, 0.105)   \\ 
 
         &  0.2      &   EigenPrism     & 0.206  & 5.86   &     ---    &   (99.0, 94.8, 88.7)    & (0.362, 0.284, 0.241)   \\ 
         &              &   Est.Equ         & 0.204  & 4.29   & 8.23     &   (99.8, 98.8, 96.6)   & (0.408, 0.326, 0.279) \\ 
         &              &   Est.Equ(N)    & 0.204  & 4.29   & 2.94     &   (96.6, 89.6, 83.1)   & (0.273, 0.210, 0.177)   \\ 
 
         &    0.5     &   EigenPrism    & 0.510  & 5.55   &     ---    &   (98.8, 95.7, 90.0)   & (0.387, 0.295, 0.247)   \\ 
         &               &   Est.Equ        & 0.509   & 5.38   & 11.1     &   (99.3, 97.0, 94.4)   & (0.511, 0.392, 0.329) \\ 
         &              &   Est.Equ(N)    & 0.509   & 5.38   & 1.94     &   (85.8, 73.5, 66.2)   & (0.226, 0.172, 0.144)   \\ 
 
         &    0.8      &   EigenPrism    & 0.801  & 2.19   &     ---    &   (99.8, 99.4, 98.6)    & (0.372, 0.292, 0.247) \\ 
         &               &   Est.Equ         & 0.800  & 1.99   & 4.46     &   (100,  99.6, 98.9)    & (0.329, 0.254, 0.213) \\ 
         &               &   Est.Equ(N)    & 0.800  & 1.99   & 0.52     &   (81.3, 67.7, 60.1)    & (0.116, 0.088, 0.074)   \\ 
  
3200  &    0.0     &   EigenPrism    & 0.059  & 6.01   &     ---    &   (97.5, 92.3, 86.3)   & (0.307, 0.236, 0.199)   \\ 
         &              &   Est.Equ         & 0.049  & 4.79   & 13.0     &   (99.5, 97.9, 94.9)   & (0.342, 0.271, 0.234) \\ 
         &              &   Est.Equ(N)    & 0.049  & 4.79   & 10.2     &   (99.3, 96.3, 92.6)   & (0.309, 0.246, 0.212)   \\ 
 
         &    0.2     &   EigenPrism    & 0.212  & 13.2   &     ---    &   (99.0, 93.8, 87.7)   & (0.481, 0.387, 0.333)   \\ 
         &              &   Est.Equ         & 0.204  & 12.8   & 16.7     &   (99.6, 97.8, 94.4)   & (0.523, 0.425, 0.369) \\ 
         &              &   Est.Equ(N)    & 0.204  & 12.8   & 10.4     &   (98.8, 95.0, 84.8)   & (0.443, 0.355, 0.305)   \\ 
 
         &    0.5      &   EigenPrism    & 0.507  & 13.3   &     ---    &   (98.6, 94.7, 89.3)   & (0.574, 0.438, 0.368)   \\ 
         &               &   Est.Equ        & 0.502   & 13.5   & 20.2     &   (99.5, 97.6, 94.4)   & (0.687, 0.545, 0.452) \\ 
         &              &   Est.Equ(N)    & 0.502   & 13.5   & 8.92     &   (96.2, 89.4, 83.1)   & (0.484, 0.369, 0.310)   \\ 
  
         &    0.8      &   EigenPrism    & 0.801  & 8.04   &     ---    &   (99.7, 97.8, 96.9)    & (0.479, 0.392, 0.340)  \\ 
         &               &   Est.Equ         & 0.800  & 7.72   & 12.0     &   (99.9, 98.1, 95.5)    & (0.471, 0.385, 0.333) \\ 
         &               &   Est.Equ(N)    & 0.800  & 7.72   & 5.51     &   (96.3, 90.5, 83.9)    & (0.359, 0.281, 0.238)   \\ 
\hline\hline
\end{tabular}
%\end{center}
\label{Sim8hnon}
\end{table}

 \begin{table}[!h] 
\caption{Simulation results on the explained variation with squared normal covariates and exponentially distributed random error, and a sample size of $300$.}
%\begin{center}
\begin{tabular}{ccl lll cc} \hline \hline
         &               &                      &             &  var              &  e.var                     &  (99, 95, 90)\% & (99, 95, 90)\%  \\
    p    &  $r^{2}$ &   method       &  est.     & $\times 10^{3}$  &  $\times 10^{3}$      &   CI coverage     & CI length     \\   \hline

 300  &  0.0     &   EigenPrism    & 0.057  & 5.62   &     ---    &    (98.5, 94.6, 90.5)   &  (0.332, 0.255, 0.215)   \\ 
         &              &   Est.Equ         & 0.036  & 2.68   & 7.92     &   (99.9, 99.5, 97.1)    & (0.264, 0.209, 0.181) \\ 
         &              &   Est.Equ(N)    & 0.022  & 2.68   & 7.23     &   (99.8, 99.1, 96.7)    & (0.254, 0.202, 0.174)   \\ 
 
         &  0.2    &   EigenPrism     & 0.209  & 12.5   &     ---    &   (99.4, 96.0, 91.9)    & (0.513, 0.418, 0.359)   \\ 
         &              &   Est.Equ         & 0.201  & 9.16   & 10.2     &   (99.6, 97.6, 91.5)   & (0.443, 0.357, 0.307) \\ 
         &              &   Est.Equ(N)    & 0.201  & 9.16   & 7.89     &   (99.3, 93.3, 86.3)   & (0.402, 0.321, 0.274)   \\ 
 
         &    0.5    &   EigenPrism     & 0.508  &  8.29   &     ---    &   (  100, 99.3, 97.4)   & (0.630, 0.480, 0.403)   \\ 
         &               &   Est.Equ        & 0.504   & 7.56   & 9.47     &   (99.5, 96.4, 92.0)   & (0.492, 0.375, 0.315) \\ 
         &               &   Est.Equ(N)   & 0.504   & 7.56   & 5.27     &   (96.2, 89.1, 81.2)   & (0.372, 0283, 0.237)   \\ 

         &    0.8    &   EigenPrism    & 0.798  & 2.76   &     ---    &   (  100,   100, 99.9)    & (0.517, 0.434, 0.382) \\ 
         &               &   Est.Equ         & 0.798  & 2.44   & 4.66     &   (  100,  99.5, 97.8)    & (0.338, 0.262, 0.221) \\ 
         &               &   Est.Equ(N)    & 0.798  & 2.44   & 1.41     &   (94.1, 85.1, 77.1)    & (0.190, 0.144, 0.121)   \\ 
 
1200 &    0.0     &   EigenPrism    & 0.090  & 13.0   &     ---    &   (98.4, 94.9, 90.4)   & (0.502, 0.388, 0.328)   \\ 
         &              &   Est.Equ         & 0.069  & 10.4   & 30.2     &   (99.7, 98.0, 95.5)   & (0.516, 0.408, 0.351) \\ 
         &              &   Est.Equ(N)    & 0.069  & 10.4   & 27.6     &   (99.6, 97.4, 94.8)   & (0.496, 0.393, 0.338)   \\

         &    0.2    &   EigenPrism    & 0.230  & 25.4   &     ---    &   (99.0, 95.3, 90.7)   & (0.680, 0.553, 0.481)   \\ 
         &              &   Est.Equ         & 0.209  & 25.1   & 31.3     &   (99.4, 97.7, 94.5)   & (0.655, 0.534, 0.464) \\ 
         &              &   Est.Equ(N)    & 0.209  & 12.6   & 27.6     &   (99.4, 97.0, 93.4)   & (0.626, 0.508, 0.441)   \\ 
 
         &    0.5   &   EigenPrism    & 0.512  & 27.4   &     ---    &   (99.6, 97.4, 93.3)   & (0.832, 0.677, 0.580)   \\ 
         &              &   Est.Equ        & 0.501   & 29.0   & 35.7     &   (99.3, 95.5, 89.7)   & (0.787, 0.633, 0.541) \\ 
         &              &   Est.Equ(N)    & 0.501   & 29.0   & 23.6     &   (97.9, 92.6, 85.9)   & (0.730, 0.578, 0.491)   \\ 
 
         &    0.8     &   EigenPrism    & 0.794  & 17.3    &     ---    &   (99.9, 98.7, 96.6)    & (0.668, 0.548, 0.480)  \\ 
         &               &   Est.Equ         & 0.790  & 17.5   & 19.7     &   (100, 99.6, 93.2)    & (0.564, 0.465, 0.405) \\ 
         &               &   Est.Equ(N)    & 0.790  & 17.5   & 14.7     &   (100, 90.6, 85.1)    & (0.510, 0.418, 0.360)   \\ 
\hline\hline
\end{tabular}
%\end{center}
\label{Sim3hsupp}
\end{table}
 
\begin{table}[!h] 
\caption{Simulation results on the explained variation with squared normal covariates and exponentially distributed random error, and a sample size of $300$.}
%\begin{center}
\begin{tabular}{ccl lll cc} \hline \hline
         &               &                      &             &  var              &  e.var                     &  (99, 95, 90)\% & (99, 95, 90)\%  \\
    p    &  $r^{2}$ &   method       &  est.     & $\times 10^{3}$  &  $\times 10^{3}$      &   CI coverage     & CI length     \\   \hline

 300  &  0.0     &   EigenPrism    & 0.058  & 5.73   &     ---    &    (98.8, 94.9, 90.8)   &  (0.333, 0.257, 0.216)   \\ 
         &              &   Est.Equ         & 0.037  & 2.90   & 7.96     &   (99.4, 97.2, 94.1)    & (0.253, 0.200, 0.173) \\ 
         &              &   Est.Equ(N)    & 0.037  & 2.90   & 7.04     &   (99.9, 98.4, 96.0)    & (0.266, 0.211, 0.182)   \\ 
 
         &  0.2    &   EigenPrism     & 0.204  & 12.1   &     ---    &   (99.7, 97.1, 92.0)    & (0.509, 0.413, 0.358)   \\ 
         &              &   Est.Equ         & 0.195  & 9.04   & 9.41     &   (99.8, 93.1, 89.1)   & (0.429, 0.346, 0.298) \\ 
         &              &   Est.Equ(N)    & 0.195  & 9.04   & 7.60     &   (98.8, 91.3, 85.5)   & (0.393, 0.314, 0.270)   \\ 
 
         &    0.5    &   EigenPrism   & 0.510  &  7,74   &     ---    &   (99.9, 99.2, 98.0)   & (0.630, 0.480, 0.403)   \\ 
         &               &   Est.Equ        & 0.503   & 7.65   & 8.16     &   (98.8, 95.2, 90.6)   & (0.462, 0.351, 0.295) \\ 
         &               &   Est.Equ(N)   & 0.503   & 7.65   & 5.76     &   (96.7, 90.7, 82.6)   & (0.390, 0.297, 0.249)   \\ 

         &    0.8    &   EigenPrism    & 0.797  & 2.80   &     ---    &   (  100,   100,   100)    & (0.518, 0.435, 0.383) \\ 
         &               &   Est.Equ         & 0.796  & 3.87   & 4.99     &   (99.6,  97.4, 94.1)    & (0.347, 0.271, 0.229) \\ 
         &               &   Est.Equ(N)    & 0.796  & 3.01   & 1.41     &   (98.3, 91.6, 85.1)    & (0.280, 0.214, 0.180)   \\ 
 
1200 &    0.0   &   EigenPrism    & 0.090  & 12.9   &     ---    &   (99.2, 94.9, 88.9)   & (0.496, 0.383, 0.323)   \\ 
         &              &   Est.Equ         & 0.070  & 10.3   & 30.2     &   (99.9, 98.2, 95.6)   & (0.518, 0.410, 0.354) \\ 
         &              &   Est.Equ(N)    & 0.070  & 10.3   & 27.6      &   (99.5, 97.0, 93.6)   & (0.492, 0.389, 0.335)   \\ 

         &    0.2    &   EigenPrism    & 0.239  & 24.7   &     ---    &   (99.6, 95.9, 92.6)   & (0.834, 0.677, 0.580)   \\ 
         &              &   Est.Equ         & 0.218  & 23.4   & 30.1     &   (97.9, 94.1, 89.1)   & (0.774, 0.61.9, 0.527) \\ 
         &              &   Est.Equ(N)    & 0.218  & 23.4   & 26.6     &   (96.6, 92.4, 87.1)   & (0.739, 0.585, 0.497) \\ 
 
         &    0.5   &   EigenPrism    & 0.506  & 27.5   &     ---    &   (99.6, 97.4, 93.3)   & (0.832, 0.677, 0.580)   \\ 
         &              &   Est.Equ        & 0.495   & 29.3   & 27.2     &   (99.3, 95.5, 89.7)   & (0.787, 0.633, 0.541) \\ 
         &              &   Est.Equ(N)    & 0.495   & 29.3   & 23.9     &   (97.9, 92.6, 85.9)   & (0.730, 0.578, 0.491)   \\ 
 
         &    0.8    &   EigenPrism    & 0.791  & 18.0    &     ---    &   (99.8, 98.7, 96.3)    & (0.670, 0.551, 0.483)  \\ 
         &               &   Est.Equ         & 0.786  & 20.1   & 21.7     &   (99.0, 97.5, 95.6)    & (0.585, 0.478, 0.415) \\ 
         &               &   Est.Equ(N)    & 0.786  & 20.1   & 19.2     &   (99.0, 97.4, 95.1)    & (0.560, 0.457, 0.396)  \\ 
\hline\hline
\end{tabular}
The EstEqu and EstEqu(N) methods used five iterations to adaptively adjust the parameter in the weighting matrix.
%\end{center}
\label{Sim3hsupp}
\end{table}

\begin{table}
\caption{Results on the explained variation with correlated covariates (positive only).}
%\begin{center}
\begin{tabular}{cl lll cc} \hline \hline
                 &                                           &             &  emp.var              &  est.var                     &  (99, 95, 90)\%      & (99, 95, 90)\%  \\
 $r^{2}$  &   method                            &  est.     & $\times 10^{3}$  &  $\times 10^{3}$      &   CI coverage          & CI length     \\   \hline
\multicolumn{7}{c}{Normal case with $ n=400, p=200$} \\
 0.210c  &   EigenPrism            & 0.206   & 4.62 &     ---  &   (99.8, 97.8, 94.4)  &  (0.360, 0.295, 0.254)   \\ 
         &   Est.Equ($W_{\lambda}$)& 0.165   & 1.89 & 3.87     &   (99.6, 97.0, 91.9)   &  (0.305, 0.241, 0.204)   \\   
         &   Est.equ($W_{*}$)      & 0.210   & 4.57 & 7.24     &   (98.4, 95.5, 89.7) &  (0.370, 0.295, 0.253)   \\               
 
 0.491c  &   EigenPrism            & 0.492   & 2.53  &     --- &   (99.5, 95.4, 89.8)  &  (0.270, 0.203, 0.170)   \\ 
         &   Est.Equ($W_{\lambda}$)& 0.432   & 2.25 & 2.03     &   (92.7, 75.4, 63.2)  &  (0.230, 0.175, 0.147)   \\ 
         &   Est.equ($W_{*}$)      & 0.495   & 2.51 & 2.94     &   (95.5, 89.2, 84.3)  &  (0.262, 0.200, 0.168)   \\ 

 0.795c  &   EigenPrism            & 0.803   & 0.53 &     ---  &   (97.4, 91.0, 84.8)  &  (0.105, 0.079, 0.066)   \\ 
         &   Est.Equ($W_{\lambda}$)& 0.759   & 0.51 & 0.47     &   (81.7, 62.0, 48.3)  &  (0.112, 0.085, 0.072)   \\ 
         &   Est.equ($W_{*}$)      & 0.804   & 0.52 & 0.43     &   (91.2, 82.4, 76.3) &  (0.100, 0.076, 0.064)   \\ 

\multicolumn{7}{c}{Non-normal case with $ n=400, p=200$} \\
0.200c   &   EigenPrism            & 0.211   & 11.3  &     --- &   (94.2, 85.0, 77.2)  &  (0.344, 0.277, 0.238)   \\ 
         &   Est.Equ($W_{\lambda}$)& 0.172   & 5.87  & 20.1    &   (100, 100, 99.8)   &  (0.517, 0.430, 0.380)   \\   
         &   Est.equ($W_{*}$)      & 0.215   & 11.1  & 48.3    &   (97.2, 94.4, 91.4) &  (0.641, 0.534, 0.468)   \\               
 
 0.502c   &   EigenPrism            & 0.514   & 10.5  &     --- &   (78.1, 64.0, 56.3)  &  (0.257, 0.194, 0.162)   \\ 
         &   Est.Equ($W_{\lambda}$)& 0.438   & 9.95  & 15.3    &   (99.0, 95.8, 92.1)  &  (0.591, 0.460, 0.388)   \\ 
         &   Est.equ($W_{*}$)      & 0.517   & 10.4  & 18.0    &   (90.1, 85.5, 81.2) &  (0.526, 0.415, 0.354)   \\ 

 0.801c   &   EigenPrism            & 0.809   & 3.42  &     --- &   (58.5, 46.8, 39.8)  &  (0.101, 0.076, 0.064)   \\ 
         &   Est.Equ($W_{\lambda}$)& 0.746   & 3.80  & 7.14    &   (99.6, 95.4, 89.9)   &  (0.399, 0.311, 0.262)   \\ 
         &   Est.equ($W_{*}$)      & 0.810   & 3.38  & 3.93    &   (79.2, 71.4, 66.3) &  (0.221, 0.171, 0.145)   \\ 

\multicolumn{7}{c}{Normal case with $ n=400, p=800$} \\
 0.201c    &  EigenPrism            & 0.055   & 5.32 &  ---   &   (95.5, 80.6, 68.4)  &  (0.396, 0.305, 0.258)   \\ 
           &  Est.Equ($W_{\lambda}$)& 0.224   & 2.70 & 5.54   &   (99.3, 97.2, 95.1)  &  (0.372, 0.289, 0.244)   \\                 
 
 0.520c   &   EigenPrism            & 0.043   & 2.43 & ---   &   (4.8,  0.0,  0.0)   &  (0.402, 0.312, 0.266)   \\ 
          &   Est.Equ($W_{\lambda}$)& 0.555   & 2.41 & 1.57  &   (84.2, 74.7, 66.8)  &  (0.201, 0.153, 0.129)   \\ 

 0.805c   &   EigenPrism            & 0.037   & 0.68 &  ---   &   (0.0, 0.0, 0.0)    &  (0.414, 0.324, 0.278)   \\ 
          &   Est.Equ($W_{\lambda}$)& 0.880   & 0.49 & 0.44   &   (19.6,8.5, 6.0)    &  (0.108, 0.082, 0.069)   \\ 
\multicolumn{7}{c}{Non-normal case with $ n=400, p=800$} \\
 0.199c    &  EigenPrism            & 0.100   & 20.4 & ---   &   (77.1, 62.7, 54.5)  &  (0.368, 0.276, 0.227)   \\ 
           &  Est.Equ($W_{\lambda}$)& 0.227   & 13.6 & 48.9  &   (99.9, 99.6, 98.9)   &  (0.750, 0.628, 0.557)   \\                 
 
 0.521c   &   EigenPrism            & 0.069   & 9.10 & ---   &   (13.4,  5.1,  3.2)   &  (0.380, 0.293, 0.248)   \\ 
          &   Est.Equ($W_{\lambda}$)& 0.551   & 16.4 & 28.2  &   (98.0, 94.0, 89.9)   &  (0.733, 0.591, 0.506)   \\ 

 0.804c   &   EigenPrism            & 0.048   & 2.54 &  ---  &   (0.0, 0.0, 0.0)    &  (0.391, 0.307, 0.263)   \\ 
          &   Est.Equ($W_{\lambda}$)& 0.879   & 6.79 & 8.51  &   (81.0, 71.9, 62.4) &  (0.323, 0.267, 0.232)   \\ 
   \hline
\end{tabular}
%\end{center}
\label{Corr}
\end{table}

\begin{table}
\caption{Results on the explained variation with correlated covariates (both positive and negative).}
%\begin{center}
\begin{tabular}{cl lll cc} \hline \hline
                 &                                           &             &  emp.var              &  est.var                     &  (99, 95, 90)\%      & (99, 95, 90)\%  \\
 $r^{2}$  &   method                            &  est.     & $\times 10^{3}$  &  $\times 10^{3}$      &   CI coverage          & CI length     \\   \hline
\multicolumn{7}{c}{Normal case with $ n=400, p=200$} \\
  0.179      &   EigenPrism                       & 0.176   & 4.56 &     ---   &   (99.9, 98.3, 94.9)  &  (0.347, 0.291, 0.253)   \\ 
                 &   Est.Equ($W_{\lambda}$)& 0.183   & 2.22 & 2.87   &   (99.3, 96.7, 91.9)   &  (0.273, 0.209, 0.176)   \\   
                 &   Est.equ($W_{*}$)           & 0.180   & 4.51  & 7.71   &   (98.1, 94.7, 90.8) &  (0.365, 0.294, 0.253)   \\               
 
  0.546      &   EigenPrism                       & 0.549   & 2.26  &     ---   &   (98.5, 93.4, 88.1)  &  (0.240, 0.181, 0.151)   \\ 
                 &   Est.Equ($W_{\lambda}$)& 0.550   & 1.59 & 1.76    &   (98.5, 94.8, 90.1)   &  (0.214, 0.163, 0.137)   \\ 
                 &   Est.equ($W_{*}$)           & 0.551  & 2.24  & 2.32   &   (95.2, 88.8, 82.4) &  (0.234, 0.178, 0.149)   \\ 

   0.800     &   EigenPrism                       & 0.804   & 0.49 &     ---   &   (97.6, 91.8, 84.5)  &  (0.104, 0.078, 0.066)   \\ 
                 &   Est.Equ($W_{\lambda}$)& 0.783   & 0.50 & 0.74   &   (99.1, 94.5, 87.6)   &  (0.137, 0.104, 0.088)   \\ 
                 &   Est.equ($W_{*}$)           & 0.805  & 0.49  & 0.45   &   (91.5, 85.0, 79.4) &  (0.101, 0.077, 0.065)   \\ 

\multicolumn{7}{c}{Non-normal case with $ n=400, p=200$} \\
 0.199       &   EigenPrism                        & 0.207   & 9.63 &     ---   &   (96.3, 88.5, 83.1)  &  (0.347, 0.281, 0.242)   \\ 
                 &   Est.Equ($W_{\lambda}$)& 0.213   & 3.84 & 13.7   &   (100, 99.6, 99.0)   &  (0.498, 0.412, 0.355)   \\   
                 &   Est.equ($W_{*}$)           & 0.211   & 9.54  & 50.4   &   (97.5, 95.2, 93.1) &  (0.650, 0.540, 0.474)   \\               
 
  0.507      &   EigenPrism                       & 0.507   & 9.83  &     ---   &   (80.1, 69.5, 61.2)  &  (0.263, 0.197, 0.165)   \\ 
                 &   Est.Equ($W_{\lambda}$)& 0.456   & 7.61 & 13.3    &   (99.5, 97.6, 94.3)   &  (0.562, 0.433, 0.365)   \\ 
                 &   Est.equ($W_{*}$)           & 0.509  & 9.73  & 22.0   &   (92.5, 86.0, 82.4) &  (0.533, 0.426, 0.366)   \\ 

   0.795     &   EigenPrism                       & 0.785   & 3.72 &     ---   &   (63.7, 50.7, 42.8)  &  (0.114, 0.086, 0.072)   \\ 
                 &   Est.Equ($W_{\lambda}$)& 0.736   & 3.42 & 7.19   &   (99.4, 96.0, 90.7)   &  (0.409, 0.316, 0.267)   \\ 
                 &   Est.equ($W_{*}$)           & 0.786  & 3.69  & 3.91   &   (87.1, 80.4, 75.0) &  (0.255, 0.198, 0.167)   \\ 

\multicolumn{7}{c}{Normal case with $ n=400, p=800$} \\
0.203        &   EigenPrism                       & 0.216   & 11.0 &     ---   &   (99.1, 97.0, 92.1)  &  (0.501, 0.405, 0.349)   \\ 
                 &   Est.Equ($W_{\lambda}$)& 0.203   & 4.98 & 5.37   &   (97.5, 93.4, 88.8)   &  (0.345, 0.270, 0.229)   \\                 
 
   0.497     &   EigenPrism                       & 0.582   & 6.60  &     ---   &   (99.7, 96.6, 90.4)  &  (0.596, 0.455, 0.382)   \\ 
                 &   Est.Equ($W_{\lambda}$)& 0.522   & 3.99 & 3.99    &   (97.4, 90.9, 85.6)   &  (0.324, 0.247, 0.207)   \\ 

  0.804     &   EigenPrism                       & 0.906   & 2.44 &     ---   &   (100, 99.2, 95.6)  &  (0.391, 0.320, 0.283)   \\ 
                 &   Est.Equ($W_{\lambda}$)& 0.806   & 1.78 & 2.08   &   (99.0, 95.1, 91.6)   &  (0.233, 0.177, 0.149)   \\ 
\multicolumn{7}{c}{Non-normal case with $ n=400, p=800$} \\
 0.198      &   EigenPrism                       & 0.098   & 10.76 &     ---   &   (91.4, 81.3, 71.7)  &  (0.387, 0.306, 0.262)   \\ 
                &   Est.Equ($W_{\lambda}$)& 0.442   & 34.4 & 84.2   &   (99.3, 95.6, 87.0)   &  (0.851, 0.720, 0.630)   \\

  0.490     &   EigenPrism                       & 0.154   & 8.02  &     ---   &   (43.2, 14.7, 6.6)  &  (0.467, 0.384, 0.336)   \\ 
                &   Est.Equ($W_{\lambda}$)& 0.622   & 11.3 & 31.4    &   (98.9, 92.7, 85.1)   &  (0.769, 0.626, 0.537)   \\

   0.806   &   EigenPrism                       & 0.238   & 2.22 &     ---   &   (0.0, 0.0, 0.0)  &  (0.556, 0.464, 0.401)   \\ 
               &   Est.Equ($W_{\lambda}$)& 0.949   & 3.55 & 19.2   &   (99.9, 94.8, 82.9)   &  (0.398, 0.315, 0.272)   \\ 
   \hline
   \hline
\end{tabular}
%\end{center}
\label{Corrpm}
\end{table}

\end{document}

%% file: estequ.tex
 \section{Introduction}

Estimation and inference on the variation explained by a set of covariates are of particular importance in scientific research,  such as the heritability in genetic studies (Yang et al, 2010) and the signal-noise ratio in wireless communications (Eldar and Chan, 2003) and in Magnetic Resonance Imaging studies (Benjamini and Yu, 2013).   When the dimension of the covariates is low, the regression coefficients may be first estimated and then plugged in the variation expression to obtain a consistent estimator. Such an approach becomes problematic when the covariate dimension is high relative to the sample size.  Under the sparse covariate effects assumption, methods for estimating the explained variation in a high-dimensional linear model have been proposed (Sun and Zhang, 2012; Fan et al., 2012; Guo and Cai, 2018; Verzelen and Gassiat, 2018; Cai and Guo, 2020 among others). However, these estimators can have very poor performance when the sparsity assumption is not true.  For example, in environmental health studies of chemical pollutants, the effects of individual pollutants are weak and difficult to identify. Collectively, the total effects of all the chemical pollutants may be significant. Our study is motivated by the need to assess the variation of a biological marker explained by all the chemical pollutants.  In such a case, a more attractive approach does not make the sparsity assumption on the covariate effects. 

When the covariate effects are dense,  Yang et al (2010) proposed a method based on the normal random effects model,  termed the genetic complex trait analysis (GCTA), for estimating the explained variation (a.k.a., the narrow-sense heritability) of a quantitative trait by the  single nucleotide ploymorphisms (SNPs) in a genome-wide association study. Their estimator is consistent when the SNPs are independent and the number of SNPs is in the comparable order of the sample size (Jiang et al, 2016).  Dicker (2014) proposed a moment-based estimator for the explained variation under the joint normality assumption of the covariates and outcome when the covariate effects are fixed instead of random. Janson et al (2017) proposed an alternative approach termed EigenPrism to construct confidence intervals for the explained variation also under the normality assumption for the independent covariates and the residual errors. One major advantage of these methods is that they do not need the sparse effects assumption. In addition, no tuning parameter is involved in the estimation approaches, potentially making inference on the estimand less involved. 

The approaches of Yang et al (2010), Dicker (2014), Dicker and Erdogdu (2016), and Janson et al (2017), all appears to rely on the normality assumptions on the data. For the environment pollutant data, such assumptions are hardly true. Although simulation results suggested that the estimation approaches are robust to the violation of the normality assumptions, the inference procedures are usually sensitive to the violation of the restrictive assumptions. For example, in an attempt to improve the inference accuracy based on the restricted maximum likelihood approach, Schweiger et al (2016) proposed a parametric bootstrap method for constructing confidence intervals which depends heavily on the normality assumptions. This approach can have poor performance when the normality assumptions do not hold as demonstrated in our simulation study. Other inference procedures all explicitly use the normality assumptions in their construction. Although the simulation results on the  EigenPrism method (Jansen et al, 2017) for $n\leq p$ suggested reasonable robustness against the normality assumptions, the least square approach suggested by Jensen et al (2017) for $n>p$ is not robust against the normality assumption on the residual error. Jiang et al (2016) showed that the GCTA estimator is consistent when the normal random effects are replaced by a mixture of normally distributed effects and zeros. In general, the consistency and the asymptotic distribution of such estimators have not been rigorously established without the normality assumptions. This makes the improvement on the robustness of the inference procedures very difficult to carry out.  To address these issues, we propose in this paper a novel closed-form estimator for the explained variation. Applying the spectral theory for the random matrix (Bai et al, 2010), we show that the proposed estimator is consistent and asymptotically normally distributed without  the normality assumption on either the covariates or the residual error. Inference procedures robust against the restrictive assumptions for the explained variation  are  proposed  based on the asymptotic analysis. 

The remainder of this paper is organized in the following way. In Section 2, the problem of estimating variation explained by covariates is formulated. A closed-form estimator for the explained variation is obtained from estimating equations.   The proposed estimator is shown to be consistent and asymptotically normally distributed in Section 3. Approximate inference procedures are also proposed based on the asymptotic analysis. Extensive simulations are conducted to evaluate the performance of the proposed approaches in comparison to the existing approaches in Section 4.  The proposed approach is applied to the inference on the explained variation of glycohemoglobin by the environmental pollutants  in a National Health and Nutrition Examination Survey(NHANES) data set in Section 5. Additional issues and further research are discussed Section 6. 

\section{Estimating equation approach to explained variation}

\subsection{Basic formulation of the problem}

Let $(Y_{i} ,X_{i1},\cdots,X_{ip}), i=1,\cdots,n$ be the observed data. Let $Y=(Y_{1},\cdots,Y_{n})^{t}$ be the vector of the observed outcomes and $X=(X_{1},\cdots, X_{p})$ be the observed covariate matrix, where $X_{j}=(X_{1j},\cdots,X_{nj})^{t}, j=1,\cdots,p$.  Assume that $(Y_{i} ,X_{i1},\cdots,X_{ip}), i=1,\cdots,n$ are i.i.d. sample from the population. Define the proportion of $Y_{i}$ variation explained by $(X_{i1},\cdots,X_{ip})$ as
$$
r^{2}=\frac{\mbox{var}\left\{E(Y_{i}\mid X_{i1},\cdots,X_{ip})\right\}}{\mbox{var}(Y_{i})}.
$$ 
Suppose that the linear model, 
\begin{equation}
Y_{i}=\beta_{0}+\beta_{1}X_{i1}+\cdots+\beta_{p}X_{ip}+\epsilon_{i},
\label{model1}
\end{equation}
holds, where $E(\epsilon_{i})=0$, $E(\epsilon_{i}^{2})=\sigma_{\epsilon}^{2}$. Let $\beta=(\beta_{1},\cdots,\beta_{p})^{t}$ and $\mbox{var}(X_{i1},\cdots,X_{ip})=\Sigma$. Under model (\ref{model1}), $r^{2}$ reduces to 
$$
r^{2}=\frac{\beta^{t}\Sigma\beta}{\beta^{t}\Sigma\beta+\sigma_{\epsilon}^{2}}.
$$
In the conventional setting where $p\ll n$, $r^{2}$ can be estimated by plugging in the least-square estimator for $\beta$, the empirical covariance matrix estimator for $\Sigma$,  and the residual variance estimator for $\sigma_{\epsilon}^{2}$.  It can be seen that such an estimator of $r^{2}$ is consistent and asymptotically normally distributed under suitable regularity conditions.  On the other hand, when $p$ is comparable to $n$ or $p>n$, neither the estimator of $\Sigma$ nor the estimators of $\beta$ and $\sigma_{\epsilon}^{2}$ behave well so that the plug-in estimator may no longer be consistent for $r^{2}$.

When $\Sigma$ is known, we may apply the following transformation  to decorrelate the covariates,
$$
(Z_{i1},\cdots,Z_{ip})^{t}=\Sigma^{-1/2}(X_{i1},\cdots,X_{ip})^{t}.
$$
Let $\alpha=(\alpha_{1},\cdots,\alpha_{p})^{t}=\Sigma^{1/2}\beta$. Model (\ref{model1}) for the transformed covariates can be rewritten as
\begin{equation}
Y_{i}=\beta_{0}+\alpha_{1}Z_{i1}+\cdots+\alpha_{p}Z_{ip}+\epsilon_{i},
\label{model2}
\end{equation}
for $i=1,\cdots,n$, where $\mbox{var}(Z_{i1},\cdots,Z_{ip})=I_{p}$.  It can be seen that
$ 
\sum_{k=1}^{p}\alpha_{k}^{2}=\beta^{t}\Sigma\beta
$
and
$$
r^{2}=\frac{\sum_{k=1}^{p}\alpha_{k}^{2}}{\sum_{k=1}^{p}\alpha_{k}^{2}+\sigma_{\epsilon}^{2}},
$$
which is also a function of the signal-to-noise ratio $\sum_{k=1}^{p}\alpha_{j}^{2}/\sigma_{\epsilon}^{2}$.

The GCTA approach (Yang et al, 2010) assumes that $\alpha_{1},\cdots, \alpha_{p}$ are independent identically distributed random effects with $\alpha_{k}\sim N(0,\sigma_{s}^{2}/p)$.
It estimates 
$
r^{2}=\sigma_{s}^{2}/(\sigma_{s}^{2}+\sigma_{\epsilon}^{2})
$
using the restricted maximum likelihood (REML) approach.  Both Dicker (2014) and Janson et al (2017) assumed $\alpha_{i}, i=1,\cdots,p$ as fixed effects, estimation of $r^{2}$ are derived by assuming covariates $Z$ as independent random variables. This latter treatment is more general as it also accommodates the case of random effects. The inference procedure for the GCTA estimator based on the REML can be invalidate when the normally distributed random effects assumption is violated. The parametric bootstrapping proposed by Schweiger et al (2015) for the GCTA estimator draws heavily on the normality assumptions and is not robust to such assumptions. The inference procedures of Dicker (2014) relies on the joint normal assumption of the outcome and covariates. The EigenPrism approach (Janson et al., 2017)  seeks optimal weights for weighting the estimating equations obtained from the singular value decomposition. The construction of the weighted estimating equation and their variance approximation also relies on the normality assumption for both the covariate $Z$ and the residual error $\epsilon$. In general, the accuracy of these inference procedures can be substantially compromised if the normality assumptions do not hold.  We propose a new estimation and inference approach in the next section that does not rely on these assumptions. 

\subsection{Estimating equation approach}

 Consider two different formulations of the problem in the literature. The first assumes a random effects model where $\alpha_{j}, j=1,\cdots,p$ are independent random effects with mean $0$ and variance $\sigma_{s}^{2}/p$ (Yang et al, 2010; Jiang et al, 2016; Schweiger et al; 2016). Note that
\begin{equation}
Y_{i}-\bar{Y}=\sum_{j=1}^{p}(Z_{ij}-\bar{Z}_{+j})\alpha_{j}+\epsilon_{i}-\bar{\epsilon},
\label{eq1}
\end{equation}
where $\bar{Y}=\sum_{i=1}^{n}Y_{i}/n$, $\bar{Z}_{+j}=\sum_{i=1}^{n}Z_{ij}/n$, and $\bar{\epsilon}=\sum_{i=1}^{n}\epsilon_{i}/n$. It follows from (\ref{eq1}) that, conditional on covariates,  
\begin{equation}
\mbox{Cov}(Y_{i}-\bar{Y},Y_{k}-\bar{Y})=\frac{1}{p}\sum_{j=1}^{p}(Z_{ij}-\bar{Z}_{+j})(Z_{kj}-\bar{Z}_{+j})\sigma_{s}^{2}+(\delta_{ik}-1/n)\sigma_{\epsilon}^{2},
\label{eq2}
\end{equation}
where $\delta_{ik}=1$ if $i=k$, and $0$ otherwise. The second formulation assumes that $Z_{j}, j=1,\cdots, p$ are independent random covariates (Dicker, 2013; Dicker and Erdogdu, 2016; Janson et al, 2017; Verzelen and Gassiat, 2018). It follows  from (\ref{eq1}) that
\begin{equation}
\mbox{Cov}(Y_{i}-\bar{Y},Y_{k}-\bar{Y})=E\left\{\sum_{j=1}^{p}(Z_{ij}-\bar{Z}_{+j})\alpha_{j}, \sum_{j=1}^{p}(Z_{kj}-\bar{Z}_{+j})\alpha_{j}\right\}+(\delta_{ik}-1/n)\sigma_{\epsilon}^{2}.
\label{eq3}
\end{equation}
Under the independence of $Z_{ij}, i=1,\cdots, n; j=1,\cdots,p$, this may be further rewritten as
\begin{equation}
\mbox{Cov}(Y_{i}-\bar{Y},Y_{k}-\bar{Y})=E\left\{\frac{1}{p}\sum_{j=1}^{p}(Z_{ij}-\bar{Z}_{+j})(Z_{kj}-\bar{Z}_{+j})\right\}\sigma_{s}^{2}+(\delta_{ik}-1/n)\sigma_{\epsilon}^{2},
\label{eq4}
\end{equation}
where $\sigma_{s}^{2}=\sum_{j=1}^{p}\alpha_{j}^{2}$. In either case, 
$$
(Y_{i}-\bar{Y})(Y_{k}-\bar{Y})-\frac{1}{p}\sum_{j=1}^{p}(Z_{ij}-\bar{Z}_{+j})(Z_{kj}-\bar{Z}_{+j})\sigma_{s}^{2}-(\delta_{ik}-1/n)\sigma_{\epsilon}^{2},
$$
for $i,k=1,\cdots,n$, are unbiased estimating scores for $\sigma_{s}^{2}$ and $\sigma_{\epsilon}^{2}$.  These scores can be rewritten in matrix form as
$$
(Y-\1_{n}\bar{Y})(Y-\1_{n}\bar{Y})^{t}-M\sigma_{s}^{2}-(I-\1_{n}\1_{n}^{t}/n)\sigma_{\epsilon}^{2},
$$
where $\1_{n}=(1,\cdots,1)^{t}$ is of length $n$ and $M$ is a $n\times n$ matrix with the entry at $i$th row and $k$th column being
$$
M_{ik}=\frac{1}{p}\sum_{j=1}^{p}(Z_{ij}-\bar{Z}_{+j})(Z_{kj}-\bar{Z}_{+j}).
$$
Let $Z=(Z_{ij})_{n\times p}$ and $\bar{Z}=(\bar{Z}_{+1},\cdots, \bar{Z}_{+p})^{t}=Z^{t}\1_{n}/n$.  It follows that 
$$
M=\frac{1}{p}(Z-\1_{n}\bar{Z}^{t})(Z-\1_{n}\bar{Z}^{t})^{t}.
$$
Let $\sigma_{Y}^{2}=\mbox{var}(Y_{1})=\sigma_{s}^{2}+\sigma_{\epsilon}^{2}$. The estimating scores can be rewritten as
$$
\frac{1}{\sigma_{Y}^{2}}(Y-\1_{n}\bar{Y})(Y-\1_{n}\bar{Y})^{t}-(I-\1_{n}\1_{n}^{t}/n)-\{M-(I-\1_{n}\1_{n}^{t}/n)\}r^{2}. 
$$
 
Let $W$ be an $n\times n$ matrix. We can obtain an estimating equation by relpacing $\sigma_{Y}^{2}$ with an empirical estimator $\hat{\sigma}_{Y}^{2}$  and weighting the estimating scores by $W$, i.e.,
$$
tr\left(W\left[\frac{1}{\hat{\sigma}_{Y}^{2}}(Y-\1_{n}\bar{Y})(Y-\1_{n}\bar{Y})^{t}-(I-\1_{n}\1_{n}^{t}/n)-\{M-(I-\1_{n}\1_{n}^{t}/n)\}r^{2}\right]\right)=0.
$$
where $tr(A)$ denotes the trace of the matrix $A$. A closed-form estimator for $r^{2}$ can be obtained as
\begin{equation}
\hat{r}^{2}=\frac{tr\left[W\left\{\frac{1}{\hat{\sigma}_{Y}^{2}}\tilde{Y}\tilde{Y}^{t}-(I-\1_{n}\1_{n}^{t}/n)\right\}\right]}
{tr\left[W\left\{M-(I-\1_{n}\1_{n}^{t}/n)\right\}\right]},
\label{direct}
\end{equation}
where $\tilde{Y}=(Y-\1_{n}\bar{Y})$. Once an estimate of $r^{2}$ is obtained,  $\sigma_{s}^{2}$ and $\sigma_{\epsilon}^{2}$ can be respectively estimated by $\hat{\sigma}_{s}^{2}=\hat{r}^{2}\hat{\sigma}_{Y}^{2}$ and $\hat{\sigma}_{\epsilon}^{2}=(1-\hat{r}^{2})\hat{\sigma}_{Y}^{2}$. 

The weighting matrix can be taken as
\begin{equation}
W_{\lambda}=(I+\lambda M)^{-1}(M-I)(I+\lambda M)^{-1},
\label{weight}
\end{equation}
where $\lambda\geq 0$ is an arbitrarily fixed constant. Denote the estimator by $\hat{r}_{\lambda}$.  Note that $\lambda$ is not a tuning parameter as any fixed $\lambda\geq 0$ can yield a consistent estimator of $r^{2}$ as shown in Theorem 1 in the next section. Different $\lambda$ values only affect the efficiency of the estimators. In the normal random effects model, an optimal choice is  $\lambda=r^{2}/(1-r^{2})$. When $n>\mbox{rank}(Z)$, the ordinary least square estimator of $\sigma_{\epsilon}^{2}$, and thus $r^{2}$,  is equivalent to taking  the weight matrix as
$
W_{*}=I-Z(Z^{t}Z)^{-}Z^{t}
$
in (\ref{direct}).  Denote the corresponding estimator by $\hat{r}_{*}$

\section{Inference on the explained variation}

\subsection{Asymptotic properties of the proposed estimators}

Since standardization of $Z$ by subtracting marginal means and divided by the marginal standard deviation does not change the essence of the  problem (Bai et al, 2010, page 503), we assume $Z$ has mean $0$ and unit variance for simplicity. Without loss of generality, we also assume $Y$ has unit variance satisfying
$$
Y=Z_{1}\alpha_{1}+\cdots+Z_{p}\alpha_{p}+\epsilon.
$$

Consider first the case with weight $W_{\lambda}$ for a fixed $\lambda$, i.e.,
\begin{equation}
\hat{r}_{\lambda}^{2}=\frac{tr\left\{W_{\lambda}(YY^{t}-I)\right\}}
{tr\left\{W_{\lambda}(M-I)\right\}},
\label{rest}
\end{equation}
with $M=ZZ^{t}/p$ and $W_{\lambda}=(I+\lambda M)^{-1}(M-I)(I+\lambda M)^{-1}$. To study the asymptotic behavior of the proposed estimator, assume that the observed data $(Y_{i},Z_{i1},\cdots,Z_{ip}), i=1,\cdots,n$ are independent identically distributed copies of $(Y, Z_{1},\cdots,Z_{p})$. In addition, we make the following assumptions.
\begin{enumerate}
\item The covariates $Z_{1},\cdots,Z_{p}$ are independent identically distributed random variables with $E(Z_{k})=0$, $\mbox{var}(Z_{k})=1$, and have the eighth finite moments.
\item The random error $\epsilon$ is independent of $(Z_{1},\cdots,Z_{p})$ with mean $0$, variance $\sigma_{\epsilon}^{2}>0$, and have the eighth finite moments.
\item As $n\rightarrow \infty$, $n/p\rightarrow \xi$, where $0<\xi<\infty$. $\sigma_{s}^{2}=\sum_{k=1}^{p}\alpha_{k}^{2}=1-\sigma_{\epsilon}^{2}$, where $\alpha=(\alpha_{1},\cdots,\alpha_{p})$ are fixed. 
\item Either $E(Z_{ij}^{4})=3$ for $i=1,\cdots,n,  j=1,\cdots, p$ or $\max_{1\leq k\leq p}|\alpha_{k}|\rightarrow 0$ as $n\rightarrow \infty$. 
\end{enumerate}

\vspace{0.3cm}

\noindent {\em Theorem 1}. Under conditions 1-3, for any fixed $0\leq \lambda<\infty$, the explained variation estimator in (\ref{rest}) is consistent, i.e.,  $\hat{r}_{\lambda}^{2}\rightarrow r^{2}$ in probability.  

\vspace{0.3cm}

In comparison to existing results in the literature, Theorem 1 states that consistency of the estimator does not require the normal distribution of either the covariates or the random error provided that the covariates are independent themselves and independent of the random error. Consistency of both $\sigma_{s}^{2}$ and $\sigma_{\epsilon}^{2}$ estimators can be derived accordingly. 

\vspace{0.3cm}

\noindent {\em Theorem 2}. Under conditions 1-4, for any fixed $0\leq\lambda<\infty$, the explained variation estimator is asymptotically normally distributed, i.e., $\sqrt{n}(\hat{r}_{\lambda}^{2}-r^{2})\rightarrow N(0,v_{\lambda}^{2})$ in distribution, where 
$$
v_{\lambda}^{2}=\frac{1}{C^{2}}\left\{2\tau^{2}\sigma_{s}^{4}/\xi+4\sigma_{\epsilon}^{2}\sigma_{s}^{2}S_{1}+2\sigma_{\epsilon}^{4}S +T (E\epsilon^{4}-3\sigma_{\epsilon}^{4})\right\},
$$
with $S=\lim _{n\rightarrow\infty}n^{-1}tr(W_{\lambda}^{2})$,
$S_{1}=n^{-1}\lim_{n\rightarrow \infty}tr(W_{\lambda}^{t}W_{\lambda}M)$, 
$T=\lim _{n\rightarrow\infty}n^{-1}\sum_{i=1}^{n}W_{\lambda ii}^{2}$, 
 and $C=\lim_{n\rightarrow \infty}n^{-1}tr\left\{W_{\lambda}(M-I)\right\}$, and
$$
\tau^{2}=\int_{(1-\sqrt{\xi})^{2}}^{(1+\sqrt{\xi})^{2}} \left\{\frac{x(x-1)}{(1+\lambda x)^{2}}\right\}^{2}d\nu(x)-\left\{\int_{(1-\sqrt{\xi})^{2}}^{(1+\sqrt{\xi})^{2}} \frac{x(x-1)}{(1+\lambda x)^{2}}d\nu(x)\right\}^{2},
$$
and $\nu$ is the limiting spectral measure in the Marchenko-Pastur law (see the Supplementary Material for more details).

\vspace{0.3cm}

When $Z_{ij}$ is normally distributed, Condition 4 is satisfied because $E(Z_{ij}^{4})=3$ (Silverstein, 1989). When $Z_{ij}$ is not normally distributed,  Condition 4 is still satisfied as long as $\max_{1\leq k\leq p}|\alpha_{k}|\rightarrow 0$ as $n\rightarrow \infty$ (Theorem 1.3 of Pan and Zhou, 2008). The latter assumption appears quite reasonable in the context of dense weak effects, i.e., nonzero coefficients are small and dense. Note also that,  in a high-dimensional setting, $\alpha$ can have a fixed nonzero $l^{2}$-norm (explained variation) even if the maximum norm of $\alpha$ goes to zero.
 
\vspace{0.3cm}

Consider next the case with weight $W_{*}$. Because $W_{*}Z=0$, it follows that
$$
\hat{r}_{*}^{2}=1-\frac{tr(W_{*}YY^{t})}{tr(W_{*})}%=1-\frac{1}{n-p}Y^{t}\left\{I-Z(Z^{t}Z)^{-1}Z^{t}\right\}Y.
$$
When $\xi>1$, $Z^{t}Z$ is invertible almost sure.  In this case, it follows from $Z=X\Sigma^{-1/2}$ that
$$
W_{*}=I-Z(Z^{t}Z)^{-1}Z^{t}=I-X(X^{t}X)^{-1}X^{t}.
$$
This means $r^{2}$ can be estimated without requiring the covariance matrix $\Sigma$ is known. 

\vspace{0.3cm}

\noindent {\em Theorem 3}. Under conditions 2-3, if $\xi>1$, the explained variation estimator $\hat{r}_{*}^{2}$ is consistent and asymptotically normally distributed, i.e., $\hat{r}_{*}^{2}\rightarrow r^{2}$ in probability, and $\sqrt{n}(\hat{r}_{*}^{2}-r^{2})\rightarrow N(0,v_{*}^{2})$ in distribution, where
$$
v_{*}^{2}=\frac{2\sigma_{\epsilon}^{4}(1-1/\xi)+T_{*}(E\epsilon^{4}-3\sigma_{\epsilon}^{4})}{(1-1/\xi)^{2}}.
$$
with $T_{*}=\lim _{n\rightarrow\infty}n^{-1}\sum_{i=1}^{n}W_{* ii}^{2}$.

\vspace{0.3cm}

The condition $\xi>1$ means the sample size can be comparable in the order to,  but reasonably larger than, the dimension of covariates. When $\xi\leq 1$ and the covariates are correlated with unknown $\Sigma$, consistent estimation of the explained variation can be unattainable (Verzelen and Gassiat, 2018). We address this issue separately when supplementary covariate data are available. Proofs of Theorems 1-3  are given in the Supplementary Material. 

\subsection{Variance estimates for approximate inference}

When it is known that $\alpha=0$, such as in hypothesis testing for $\sigma_{s}^{2}=0$,  the numerator in the $v_{\lambda}^{2}$ expression can be approximated by
$$
\frac{2\sigma_{\epsilon}^{4}}{n}tr(W_{\lambda}^{t}W_{\lambda})+\frac{1}{n}\sum_{i=1}^{n}W_{\lambda ii}^{2}\left(E\epsilon^{4}-3\sigma_{\epsilon}^{4}\right),
$$
which can be consistently estimated by
$$
\frac{2}{n}\left\{tr(W_{\lambda}^{t}W_{\lambda})-\sum_{i=1}^{n}W_{\lambda ii}^{2}\right\}+\frac{1}{n}\sum_{i=1}^{n}W_{\lambda ii}^{2}(Y_{i}^{2}-1)^{2}.
$$
The variance $v_{\lambda}^{2}$ can be consistently estimated by
$$
\hat{v}_{\lambda}=\frac{\frac{2}{n}\left\{tr(W_{\lambda}^{t}W_{\lambda})-\sum_{i=1}^{n}W_{\lambda ii}^{2}\right\}+\frac{1}{n}\sum_{i=1}^{n}W_{\lambda ii}^{2}(Y_{i}^{2}-1)^{2}}{[tr\{W_{\lambda}(M-I)\}/n]^{2}}.
$$
When $\alpha\neq 0$ or it is not known that $\alpha=0$, if the random error is normally distributed, the variance $v_{\lambda}^{2}$ can be consistently estimated by
$$
\tilde{v}_{\lambda}^{2}=\frac{2\hat{r}_{\lambda}^{4}\hat{\tau}^{2}p/n+4\hat{r}_{\lambda}^{2}(1-\hat{r}_{\lambda}^{2})tr(W_{\lambda}^{t}W_{\lambda}M)/n+2(1-\hat{r}_{\lambda}^{2})^{2}tr(W_{\lambda}^{t}W_{\lambda})/n}{[tr\{W_{\lambda}(M-I)\}/n]^{2}},
$$
where $p\wedge n=\min(p,n)$, 
$$
\hat{\tau}^{2}=\frac{1}{p}\sum_{k=1}^{p\wedge n}\left\{\frac{\eta_{k}(\eta_{k}-1)}{(1+\lambda\eta_{k})^{2}}\right\}^{2}-\left\{\frac{1}{p}\sum_{k=1}^{p\wedge n}\frac{\eta_{k}(\eta_{k}-1)}{(1+\lambda\eta_{k})^{2}}\right\}^{2},
$$
and $\eta_{k}, k=1,\cdots, p\wedge n$ are non-zero eigenvalues of $M$. In general, a consistent estimate of the asymptotic variance $v_{\lambda}^{2}$ is 
$$
\hat{v}_{\lambda}^{2}=\tilde{v}_{\lambda}^{2}-\frac{2(1-\hat{r}_{\lambda}^{2})^{2}\frac{1}{n}\sum_{i=1}^{n}W_{\lambda ii}^{2}}{[tr\{W_{\lambda}(M-I)\}/n]^{2}}+\hat{va}1_{\{\hat{va}\geq 0\}}.
$$
where
$$
\hat{va}=\frac{\frac{1}{n}\sum_{i=1}^{n}\{Y_{i}^{2}-1-(M_{ii}-1)\hat{r}_{\lambda}^{2}\}^{2} -4\hat{r}_{\lambda}^{2}(1-\hat{r}_{\lambda}^{2})-2\hat{r}_{\lambda}^{4}}{[tr\{W_{\lambda}(M-I)\}/n]^{2}}\frac{1}{n}\sum_{i=1}^{n}W_{\lambda ii}^{2}.
$$
More details on the derivation of the variance estimate can be found in the Supplementary Material. 

The variance $v_{*}^{2}$ can be consistently estimated by 
$$
\tilde{v}_{*}^{2}=\frac{2n}{n-p}(1-\hat{r}_{*}^{2})^{2},
$$
when the random error is normally distributed. Without the normality assumption, the variance can be approximated by
$$
\hat{v}_{*}^{2}=\frac{2p(1-\hat{r}_{*}^{2})^{2}}{n-p}+\max\left\{\frac{\sum_{j=1}^{n-p}\left(U_{j}^{*t}Y\right)^{4}-3(n-p)(1-\hat{r}_{*}^{2})^{2}}{\sum_{j=1}^{n-p}\sum_{k=1}^{n}U_{jk}^{*4}}+2(1-\hat{r}_{*}^{2})^{2},0\right\},
$$ 
where $U_{(n-p)\times n}^{*}$ satisfying $U^{*t}U^{*}=I_{n-p}$ and $I-Z(Z^{t}Z)Z^{t}=U^{*}U^{*t}$, and $U^{*t}=(U_{1}^{*t},\cdots,U_{n-p}^{*t})$ and  $U_{j}^{*}=(U_{j1}^{*},\cdots,U_{jn}^{*})$. See the Supplementary Material for the more detailed derivation.
 
We propose to make inference on $r^{2}$ based on the normal approximation to the asymptotic distribution of $\hat{r}_{\lambda}^{2}$ or $\hat{r}_{*}^{2}$ with the respective variance estimates.  Inference on parameter $\sigma_{s}^{2}$ and $\sigma_{\epsilon}^{2}$ can be carried out accordingly following the $\delta$-method for obtaining their asymptotic distributions respectively.

\section{Simulation study}

 Simulations are conducted in this section to study the performance of the proposed approach in comparison with both the EigenPrism approach and  the GCTA approach. Since the EigenPrism approach in Janson et al (2017) is developed for $p\geq n$. For the case $p<n$, they recommended using the least square estimator for $\sigma_{\epsilon}^{2}$ and set $\sigma_{s}^{2}=\mbox{var}(Y_{1})-\sigma_{\epsilon}^{2}$. In the simulation study when $n>p$, the EigenPrism approach computes the $100(1-\alpha)\%$ confidence interval by 
$$
P\left\{\chi_{n-p-1}^{2}(\alpha/2)\leq (n-p-1)\frac{1-\hat{r}_{*}^{2}}{1-r^{2}}\leq \chi_{n-p-1}^{2}(1-\alpha/2)\right\}=1-\alpha,
$$
where $\hat{r}^{2}$ is the estimator corresponding to using $W^{*}$ as the weight matrix.  This leads to the confidence interval 
$$
1-\frac{1-\hat{r}_{*}^{2}}{\chi_{n-p-1}^{2}(\alpha/2)/(n-p-1)}\leq r^{2}\leq 1-\frac{1-\hat{r}_{*}^{2}}{\chi_{n-p-1}^{2}(1-\alpha/2)/(n-p-1)}.
$$
The GCTA approach corresponds to use $W_{\lambda}$ with $\lambda$ estimated adaptively. For inference, we adopt the parametric bootstrapping idea proposed in Schweiger et al (2016) in estimating the distribution of the estimator in the GCTA approach. For the proposed approach, we start $\lambda=0.1$ with $5$ iterations using $\lambda=\hat{r}^{2}/(1-\hat{r}^{2})$.  Two ways of inference are used: One uses the consistent variance estimate under the normality assumption for the random error (EstEqu(N)), the other uses the estimated variance without relying on the normality assumption (EstEqu).

The first simulation study evaluates the performance of $\hat{r}_{\lambda}^{2}$. The sample sizes are set to $n=200$ or $n=800$. For each sample size, we set $p=n/2, n, 4n$ respectively. Both the covariates and the random errors are generated as independent standard normal with power transformation to represent deviation from the normal distribution. Specifically, a normal random number $u$ is transformed into $u_{1}$  through
$
u_{1}=\mbox{sign}(u)*|u|^{\gamma},
$
where $\gamma>0$, and $\gamma=1$ corresponds to normal. In the case of $\chi_{1}^{2}$ covariates, the transformation drops $\mbox{sign}(u)$. The covariate effects are set to fixed constant for half of the variables and $0$ for the other half. The constants are chosen to make the explained variations at $0, 2, 5, 8$ and the error variances are chosen to be $10, 8, 5, 2$ respectively so that the proportion of explained variation are $0, 0.2, 0.5, 0.8$ respectively. The simulation results are based on $1000$ replicates. The bootstrap sample size is set to $500$ in the inference for the GCTA approach. For each estimator, we compute the coverage rates of the  $95\%$ confidence intervals along with the average length of these confidence intervals.

Figure \ref{fig1} displays the coverage rate (curves on the top of each graph) and length (vertical lines at the bottom of each graph) of the 95\% confidence intervals for the explained variations for $n=200$. The top panel of Figure \ref{fig1} compares the methods under model (\ref{model2}) with normally distributed covariates and normally distributed random error.  The bottom panel of figure \ref{fig1} compares the methods with non-normally distributed covariates and non-normally distributed random error. The non-normal covariates are $\chi^{2}$ distributed with one degree of freedom and the non-normal random error is the transformation with $\gamma=3$.  From Figure \ref{fig1}, we see that the GCTA approach can produce confidence intervals with substantially lower coverages than the nominal levels, in particular, in the non-normal case. So is the EstEqu(N) method.  When $n\leq p$, both the EigenPrism approach and the EstEqu method produce confidence intervals having at-least the nominal coverage. The EstEqu method has shorter confidence intervals than the EigenPrism approach in the normal cases, while the EigPrism approach has shorter confidence intervals than the EstEqu in the simulated non-normal cases. When $n>p$, the EstEqu method is the only one that can produce confidence intervals having at least the nominal covarage in both normal and non-normal cases. 

The sample size is increased to $800$ and other settings remain unchanged in the second set of simulation. Since computing the bootstrap variance estimate in the GCTA approach is very time-consuming and the simulation results with the sample size of $200$ suggest its performance is close to the EstEqu(N), it is excluded in the second set of simulations. Figure \ref{fig2} displays the coverage rate and length of the 95\% confidence intervals for the explained variations. The layout of figure \ref{fig2} is the same as in figure \ref{fig1} except that the GCTA method is excluded. It can be seen from figure \ref{fig2} that the coverage rate and length of the 95\% confidence intervals show similar patterns as observed in Figure \ref{fig1} for the sample size $200$. More detailed simulation results on the bias of the estimators, the performance of the variance estimates as well as the coverage and length of the 90\% and 99\% confidence intervals under the simulated scenarios can be found in the Supplementary Material. 
 
In summary, the simulation results demonstrate that the proposed EstEqu method produces confidence intervals having good coverages for all the cases whether $n<p$ or $n\geq p$, whether normal or non-normal covariates and random errors. When $n\leq p$, the EigenPrism approach has also good coverages. The GCTA approach does not maintain good coverages. The EstEqu method has better performance in some cases while the EigenPrism approach has better performance in other cases. Although the EigenPrism approach has better performance than the EstEqu method in the non-normal cases in the simulation, we note that this is not always the case. If we replace the cubed random error with the exponential random error, the EstEqu method would have better performance in the non-normal cases. Such simulation results can be found in the Supplementary Material. The choice of constant covariate effects in the simulation is not critical either. When such effects are replaced by non-constant effects generated from a normal distribution, the comparison among the methods shows similar patterns. 

The second simulation study evaluates the impact of the correlation among covariates on the estimation and inference of the explained variation. Covariates with different levels of correlation are generated as follows. First, generate a $p\times p$ matrix $A$ of independent normal $N(a, 1)$ entries and a $p\times p$ matrix $B$ of independent uniform $U[-0.5,0.5]$ entries. Create a covariance matrix $D=(AB)^{t}(AB)$. The correlation matrix $C$ corresponding to the covariance matrix $D$ is then obtained. The parameter $a$ is used to adjust the levels of correlation. The simulations chose $a=2$ to yield correlation coefficients in $[-0.95,0.95]$ approximately. Absolute values are taken to generate a correlation matrix similar to the correlation distribution of the PCBs in the NHANES data set. See figure \ref{corrdistr}.  A correlation matrix $C$ is extracted from the covariance matrix $D$ using the singular value decomposition approach. Covariates are first generated as standard normal random numbers and the power transformation may be applied to generate deviation from the normal. The covariates are then rescaled by their standard deviation before transformed by $C^{1/2}$. The rest of data generation remains the same as in the first simulation study.  We simulated $n=400, p=200$ and $n=400, p=800$ respectively. When $n>p$,  the EigenPrism approach assumes normal error distribution but not independent covariates, the weighted estimating equation approach assumes independent covariates but not normal error distribution, and the least square approach does not assume independent covariates with or without normal error distribution. In addition, an approach termed TransEE that uses the estimated correlation matrix to decorrelate the covariates and then applies the estimating equation approach to the decorrelated covariates is also included in the comparison. When $n<p$, the least square approach cannot be applied, nor the TransEE approach. Both the EigenPrism and the weighted estimating equation approaches assume  independent covariates.  

The simulation results are listed in Table \ref{Corr}. It can be seen from the results, the proposed estimating equation approach, though subject to some bias, is fairly robust against the violation of independent covariates assumption, especially when the explained variation is not very big. As expected, the EigenPrism approach may not perform well when the normality assumptions are violated and $n>p$. It is a little surprising that the least-square approach, though asymptotically unbiased,  can have low coverage rates when $n>p$. The TransEE appears to perform the best among the methods compared  in terms of the coverage when $n>p$. For the case where $n<p$, the EigenPrism approach can have very poor performance. On the other hand, the weighted estimating equation approach appears to have relatively small bias, though the covarage rate can be low. This might be partially due to the incorrect variance estimates.  Overall, in terms of mean square error, the estimating equation assuming independence has the smallest.  In addition, correlation among covariates affects the estimation and inference of the explained variation much less when the explained variation is relatively small than it is large.

\section{Application to the NHANES data}

In this section, we apply the proposed approach to study the polychlorinated biphenyl (PCB) effects on the hemoglobin level in a data set from the national Health and Nutrition Examination Survey (NHANES). The data set contains several health outcomes as well as demographical variables, measurements of environmental pollutants such as PCBs and heavy metals. In this analysis, we concentrate on the effects of environmental PCBs on the glycohemoglobin level in human body. The glycohemoglobin is a biological measurement related to type II diabetes. The data set has 977 subjects, each with 38 PCBs measured. The major objective of this analysis is to examine the variation of the glycohemoglobin explained by the 38 PCBs. 

By examining the empirical distributions of the PCBs (the boxplots in figure \ref{boxplot}) and the correlations among the PCBs, we notice two important features of the data. First, the PCBs are highly correlated. Although this data set has a sample size larger than the number variables, the fact that the variables are highly correlated hampers the estimation of the individual variable effects, in particular, the effects of individual interactions. Second, the distributions of the PCBs are heavily tailed. The log-transformed PCBs are much closer to normally distributed.  
 
 We first estimate the hemoglobin variation explained by the PCBs in the linear model with main effects.  Five methods are applied to the data. They are the EigenPrism approach, the GCTA approach, the weighted estimating equation approach with weight $W_{\lambda}$ with or without assuming normality for the covariates and the random error, the transEE approach with weight $W_{\lambda}$ with or without assuming normality for the covariates and the random error,  and the least-square approach with or without assuming normal random error.  The results are displayed in Table \ref{DataAnalysis}. When the PCBs are in the original scale, the estimated $r^{2}$s of the main effects are around $10\sim 12\%$. The estimated $r^{2}$s of the main effects are in the range of $13\sim 15\%$ for the log-transformed PCBs. Under the normal assumption, the main effects are significantly different from $0$. Evidence for non-zero mean effects disappear without the normality assumption.  

We next estimate the hemoglobin variation explained by the PCBs with both main effects and pairwise interactions.  The results are also in Table \ref{DataAnalysis}. The estimated total explained variation is around $15\sim 40\%$ for PCBs in the original scale and $14\sim 27\%$ for the log-transformed PCBs. The $95\%$ confidence intervals for the combined explained variation are all significantly different from $0$ under the  normality assumptions.  Without the normality assumption, both the TransEE and the EstEqu($W_{*}$) yield significant effects. The method of EstEqu($W_{\lambda}$) does not yield significant effects. From the simulation experiment, both the TransEE and the EstEqu($W_{*}$) can be more reliable within the range of the estimated effects. These results suggest that the PCB pollutants interact and can significantly predict the glycohemoglobin variation. 

\section{Discussion}

We proposed an estimating equation approach to making inference on the variation explained by a set of  high-dimensional covariates in a linear model. Our simulation results demonstrate the proposed estimator has good finite-sample performance in comparison to other existing approaches. Although the EigenPrism approach has reasonably good performance in comparison to the proposed approach when covariates are independent, it is surprising that the performance of the EigenPrism approach under the correlated covariates can be much worse than the proposed approach. This might serve as a warning sign in applying the EigenPrism approach in practice.

The proposed estimator is shown to be consistent and asymptotically normally distributed. Although other estimators had been proposed, such properties have not been established without the strong normality assumption on the covariates and the random error when the covariate effects can be dense. The requirement of the identically distributed covariates can be relaxed and the consistency result may be strengthened to almost sure convergence. The asymptotic results are proved under the key assumption that the covariates are independent. When this assumption is violated, bias may occur in the explained variation estimator. However, simulation results demonstrated that the bias in the estimation may be relatively small even when the covariates are highly correlated as in the NHANES data example. The overall mean squared error of the proposed method can still be low with practical sample sizes. 

When $n<p$, the covariates are correlated, and the covariate effects are dense, it may not be possible to consistently estimate the explained variation (Verzelen and Gassiat, 2018). The difficulty  may be overcome with the help of supplementary covariate data. However, this is beyond the scope of this paper and is a topic of further research.

%Our simulation results demonstrate the proposed estimator has good performance in comparison to other existing approaches. When $p\geq n$, the proposed approach has comparable performance to the EigenPrism approach in terms of the confidence interval coverages and the confidence interval length. In some cases, the proposed approach has better performance; in other cases, the EigenPrism approach has better performance. In all the simulation results, both the proposed approach and the EigenPrism approach have adequate coverages. On the other hand, the confidence intervals constructed using the GCTA approach with bootstrapping variance estimate and the estimating equation approach utilizing the normality assumption can have a serious problem of under coverages. For the case  $p<n$, the confidence intervals based on the least-square approach as suggested in the EigenPrism paper can have substantial under-coverages. In contrast, the proposed approach also has adequate coverages when $p<n$.

\vspace{1in}
\begin{center} Acknowledgement\end{center}

This research is supported by a grant from the National Institute of Environmental Health Sciences at the National Institute of Health.

%% file: reference.tex
\noindent References 

\begin{description}
\item Bai, Z. D., Miao, B. Q., Pan, G. M. (2007). On asymptotics of eigenvectors of large sample covariance matrix. {\em Annals of Probability}, {\bf 35}, 1532-1572.

\item Bai, Z. D. and Silverstein, J. W. (1998). No eigenvalues outside the support of the limiting spectral distribution of large dimensional random matrices. {\em Annals of Probability}, {\bf 26}, 316-345. MR1617051

\item Benjamini, Y. and Yu, B. (2013). The schuffle estimator for explainable variance in fmri experiemnets. {\em The Annals of Applied Statistics}, {\bf 7}, 2007-2033.

%\item Brown, B. M. (1971). Martingale central limit theorems. {\em The Annals of Mathematical Statistics}, {\bf 42}, 59-66.

\item Cai, T. T. and Guo, Z. (2020). Semi-supervised inference for explained variance in high-dimensional regression and its applications. {\em Journal of the Royal Statistical Society, Ser. B.}, {\bf 82}, 391-419.

\item De Jong, P. (1987). A central limit theorem for generalized quadratic forms. {\em Probability Theory and Related Fields}, {\bf 75 }, 261-277. 

\item Dicker, L. H. (2014). Variance estimation in high-dimensional linear models. {\em Biometrika}, {\bf 101}, 269-284.
 
\item Dicker, L. H. and  Erdogdu, M. A. (2016). Maximum likelihood for variance estimation in high-dimensional linear
models. {\em Proc. Mach. Learn. Res.}, {\bf  51}, 159-167.% Proc. 19th Int. Conf. Artificial Intelligence and Statistics (AISTATS 2016).

\item Eldar, Y. C. and Chan, A. M. (2003). On the asymptotic performance of the decorrelator. {\em IEEE Tans. Inform. Theory}, {\bf 49}, 2309-2313. MR2004788.

\item Fan, J., Guo, S. and  Hao, N. (2012). Variance estimation using refitted cross-validation in ultrahigh-dimensional
regression. {\em J. R. Statist. Soc. B}, {\bf 74}, 37-65.

\item Guo, Z., Wang, W., Cai, T. T., Li, H. (2018). Optimal estimation of co-heritability in high-dimensional linear models. {\em Journal of American Statistical Association}, {\bf 114}, 358-369. 
 
\item Heyde, C. C. and Brown, B. M. (1970). On the departure from normality of a certain class of martingales. {\em The Annals of Mathematical Statistics}, {\bf 6}, 2161-2165.

\item Janson, L., Barber, R. F., Candes, E. (2017). EigenPrism: inference for high-dimensional signal-to-noise ratios. {\em Journal of Royal Statistical Society, Ser. B.}, {\bf 79}, 1037-1065.

\item Jiang, J., Li, C., Paul, D., Yang, C. and Zhao, H. (2016). On high-dimensional misspecified mixed model analysis in genome-wide association study. {\em Annals of Statistics}, {\bf 44}, 2127-2160.

%\item Kumar, S. K., Feldman, M. W., Rehkopf, D. H., Tuljapurkar, S. (2016). Limitations of GCTA as a solution to the missing heritability problem. {\em Proceedings of the National Academy of Sciences}, {\bf 113}, E61-70.

%\item Lazarevic, N., Barnett, A. G., Sly, P. D., Knibbs, L. D. (2019). Statistical methodology in studies of prenatal exposure to mixtures of endocrine-disruptive chemicals: A review of existing approaches and new alternatives. {\em Environmental Health Perspectives}, {\bf 127}, PMCID: PMC6752940. 

%\item  Lee, SH, Wray, NR, Goddard, ME, Visscher, PM (2011). Estimating missing heritability for disease from genome-wide association studies. {\em American Journal of Human Genetics}, {\bf 88}, 294-305.

\item Marchenko, V. and Pastur, L. (1967). Distribution of eigenvalues for some sets of random matrices. {\em Sbor. Mathematics}, {\bf 114}, 506-536.

\item Pan, G. M. and Zhou, W. (2008). Central limit theorem for signal-to-interference ratio of reduced rank linear receiver. {\em The Annals of Applied Probability}, {\bf 18}, 1232-1270.

\item Schweiger, R., Kaufman, S., Laaksonen, R., Kleber, M. E., Marz, W., Eskin, E., Rosset, S., Halperin, E. (2016). Fast and accurate construction of confidence intervals for heritability. {\em The American Journal of Human Genetics}, {\bf 98}, 1181-1192.

\item Silverstein, J. W. (1989). On the eigenvectors of large-dimensional sample covariance matrices. {\em Journal of Multivariate Analysis}, {\bf 30}, 1-16. MR1003705.

\item Sun, T, and Zhang, C. H. (2012). Scaled sparse linear regression. {\em Biometrika}, {\bf 99}, 879-898.

\item Verzelen, N. and Gassiat, E. (2018) Adaptive estimation of high-dimensional signal-to-noise ratios. {\em Bernoulli}, {\bf 24},
 3683-3710

\item Yang, JA, Benyamin, B, McEvoy, BP, Gordon, S, Henders, AK, Nyholt, DR, Madden, PA, Heath, AC, Martin, NG, Montgomery, GW, Goddard, ME, Visscher, PM (2010). Common SNPs explain a large proportion of the heritability for human height . {\em Nature Genetics}, {\bf 42}, 565-569.

\end{description}

%% file: figtab.tex
 
\begin{figure}[!h]
\centering
\includegraphics[height=7in,width=6.5in]{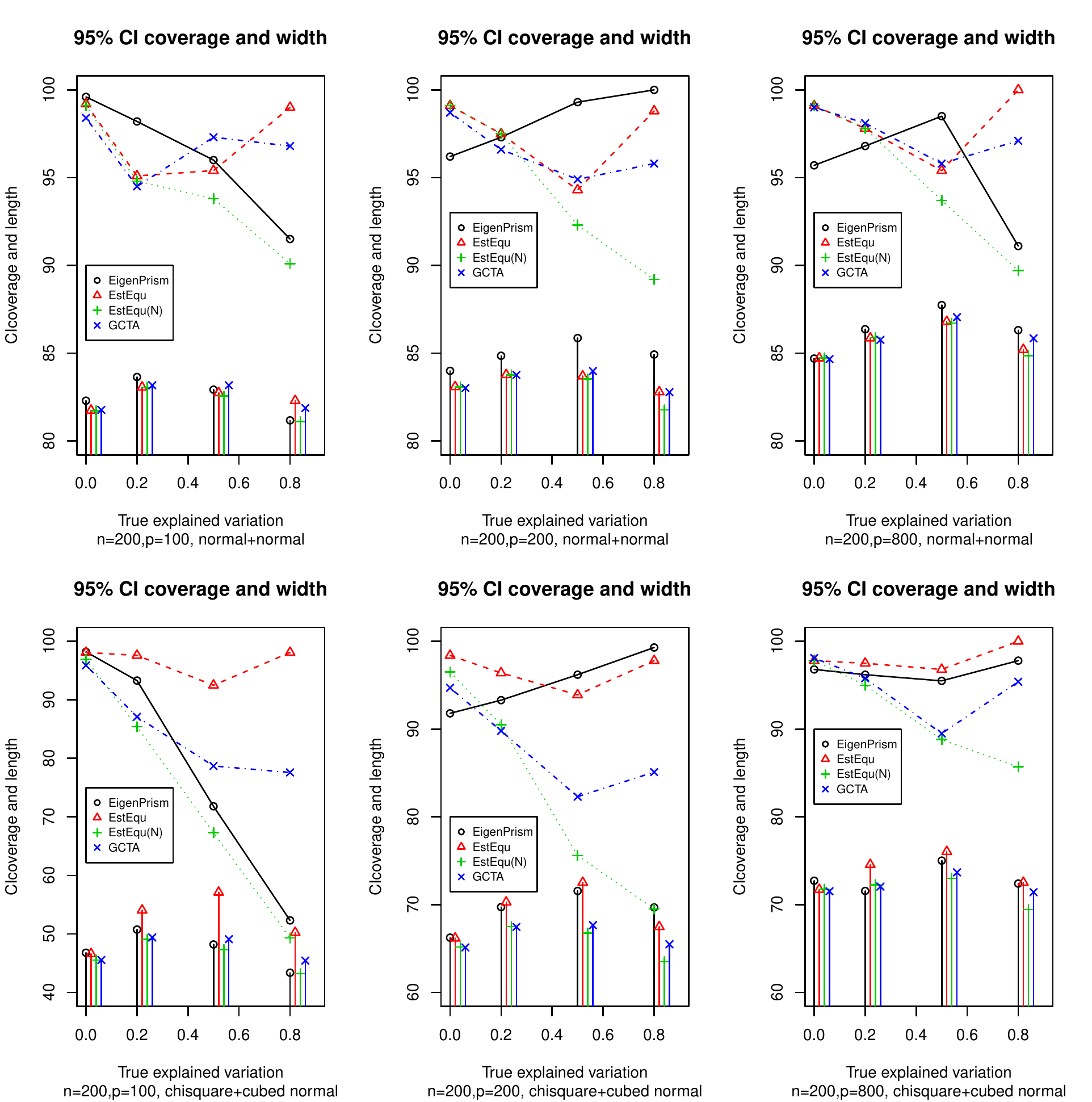}
\caption{
Coverage rates and interval length (vertical bars show the relative length) of the $95\%$ confidence intervals for the explained variation with sample size 200. %Tables 1-2 
 }
\label{fig1}
\end{figure}

\begin{figure}[!h]
\centering
\includegraphics[height=7in,width=6.5in]{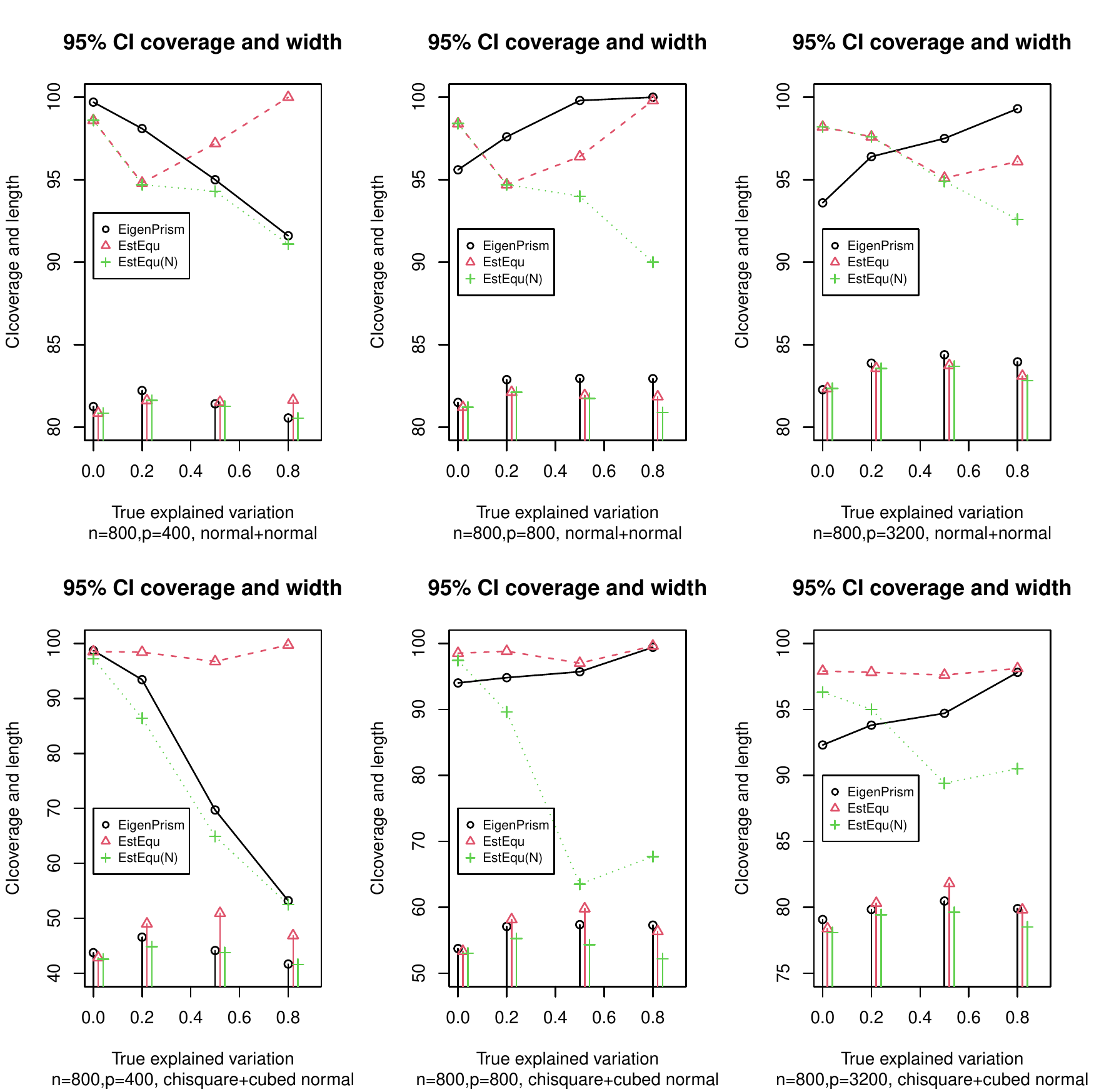}
\caption{ Coverage rates and interval length (vertical bars show the relative length) of the $95\%$ confidence intervals for the explained variation with sample size 800.%Tables 3-4 
 }
\label{fig2}
\end{figure}
 
\begin{table}[!h] 
\caption{Results on the explained variation with correlated covariates.}
%\begin{center}
\begin{tabular}{cl lll ccc ccc} \hline \hline
                 &                                           &             &                          &                   & \multicolumn{3}{c}{Assume normality} & \multicolumn{3}{c}{Not assume normalilty}  \\
                 &                                           &             &  emp.v              &  mse           &  est.v                      &   CI  &  CI &  est.v   &   CI   &  CI \\
 $r^{2}$  &   method                            &  est.     & $\times 10^{3}$  &  $\times 10^{3}$  &  $\times 10^{3}$   &   cover.  & length  &  $\times 10^{3}$   &   cover. & length \\   \hline
              & \multicolumn{9}{c}{Normal case with $ n=400, p=200$} \\
  0.211  &   EigenPrism                   & 0.208   & 4.36 &    4.37      & ---     &   98.3  &   0.296  &  ---       &  ---       &  ---      \\ 
         &   Est.Equ($W_{\lambda}$)& 0.175   & 1.92 &    3.22      & 3.46  &  98.0  &    0.228 &3.66  &   98.4  &   0.234    \\   
         &   TransEE($W_{\lambda}$)& 0.208   & 4.36 &    4.37     &  4.21  &  94.0  &   0.249 &4.45  &   94.4  &   0.256    \\   
         &   Est.equ($W_{*}$)           & 0.212   & 4.32 &    4.32     &   6.25  & 98.0  &   0.298  &7.33 &   94.3  &   0.294    \\              
       
 0.499  &   EigenPrism                   & 0.494    & 2.73 &   2.76       & ---     &   93.1  &   0.203   &  ---       &  ---       &  ---       \\ 
         &   Est.Equ($W_{\lambda}$)& 0.482   & 1.70 &   1.99       & 1.77  &  94.7  &    0.165 &0.99  &   84.0  &   0.119    \\   
         &   TransEE($W_{\lambda}$)& 0.494   & 2.73 &  2.76       &  2.20  &  92.1  &   0.183 &2.32 &   91.5  &   0.187    \\   
         &   Est.equ($W_{*}$)           & 0.496   & 2.70 &  2.71       &   2.56 & 93.3  &   0.197  &3.91 &   89.8  &   0.198    \\              

 0.791  &   EigenPrism                   & 0.799   & 0.53 &   0.59       & ---     &   90.5  &   0.081   &  ---       &  ---       &  ---     \\ 
         &   Est.Equ($W_{\lambda}$)& 0.805   & 0.40 &  0.60        & 0.28  &  78.0  &    0.065 &0.09  &   49.6  &   0.036    \\   
         &   TransEE($W_{\lambda}$)& 0.799   & 0.53 &  0.59       &  0.40  &  85.7  &   0.078 &0.51  &   84.4  &   0.082    \\   
         &   Est.equ($W_{*}$)           & 0.800   & 0.52 &  0.60       &   0.41  & 85.6  &   0.078  &0.47 &   81.4  &   0.080    \\              

         & \multicolumn{9}{c}{Non-normal case with $ n=400, p=200$} \\
0.205   &   EigenPrism                   & 0.220   & 7.58 &   7.81       & ---     &   91.4  &   0.287   &  ---       &  ---       &  ---     \\ 
         &   Est.Equ($W_{\lambda}$)& 0.185   & 3.72 &  4.12        & 3.45  &  93.6  &    0.226 &31.7  &   100  &   0.512    \\   
         &   TransEE($W_{\lambda}$)& 0.220   & 7.33 &  7.56       &  4.11  &  84.6  &   0.243 &13.4  &   96.9  &   0.409    \\   
         &   Est.equ($W_{*}$)           & 0.224   & 7.27 &  7.63       &   6.09  & 90.1  &   0.289  &42.9 &   96.5  &   0.532    \\              

 0.488   &   EigenPrism                   & 0.493   & 8.42 &  8.45        & ---     &   72.1  &   0.203   &  ---       &  ---       &  ---      \\ 
         &   Est.Equ($W_{\lambda}$)& 0.481   & 6.86 &  6.91        & 1.79  &  67.8  &    0.164 &20.8 &   95.9  &   0.498    \\   
         &   TransEE($W_{\lambda}$)& 0.493   &8.42 &   8.45      &  2.22  &  67.0  &   0.182 &14.3  &   94.6  &   0.443    \\   
         &   Est.equ($W_{*}$)           & 0.496   & 8.34 &  8.40       &  2.63  & 71.5  &   0.198  &21.8 &   89.3  &   0.449    \\              

 0.781   &   EigenPrism                   & 0.781   & 3.57 &   3.57       & ---     &   53.2  &   0.088   &  ---       &  ---       &  ---       \\ 
         &   Est.Equ($W_{\lambda}$)& 0.788   & 3.43 &   3.48        & 0.36  &  45.3  &    0.072 &3.06  &   55.9  &   0.144    \\   
         &   TransEE($W_{\lambda}$)& 0.781   & 3.57 &   3.57      &  0.50  &  51.9  &   0.084 &4.43 &   87.0  &   0.226    \\   
         &   Est.equ($W_{*}$)           & 0.782   & 3.54 &   3.54     &   0.51  & 52.3  &   0.085  &4.43 &   79.4  &   0.207    \\              

          & \multicolumn{9}{c}{Normal case with $ n=400, p=800$} \\
 0.201    &  EigenPrism                    & 0.053   & 5.28 &  27.2       & ---   &   78.4   &  0.295  & ---    & ---     &  --- \\ 
           &  Est.Equ($W_{\lambda}$)& 0.211   & 3.59 &   3.69      &5.53 &   98.5   &  0.286  & 5.85 & 98.2 & 0.293 \\                 
 
 0.494   &  EigenPrism                    & 0.045   & 3.14 &  205       & ---   &    1.5     &  0.312  & ---    & ---     &  --- \\ 
           &  Est.Equ($W_{\lambda}$)& 0.533   & 2.54 &  4.06       &2.44 &   83.5   &  0.193  & 1.53 & 71.8 & 0.150 \\    

 0.809   &  EigenPrism                    & 0.035   & 0.59 &   600      & ---    &   0.0    &  0.319  & ---    & ---     &  --- \\ 
           &  Est.Equ($W_{\lambda}$)& 0.873   & 0.54 &  4.64       &0.31 &  13.6   &  0.068  & 0.21 & 8.9 & 0.056 \\    

           & \multicolumn{9}{c}{Non-normal case with $ n=400, p=800$} \\
 0.229   &  EigenPrism                    & 0.056   & 5.77 &   35.7      & ---   &   73.2   &  0.301  & ---    & ---     &  --- \\ 
           &  Est.Equ($W_{\lambda}$)& 0.258   & 6.62 &  7.46       &5.09 &   88.6   &  0.275  & 47.6 & 99.7 & 0.653 \\    
 
 0.466   &  EigenPrism                    & 0.044   & 3.10 &  180       & ---   &   3.0    &  0.309  & ---    & ---     &  --- \\ 
           &  Est.Equ($W_{\lambda}$)& 0.537   & 9.26 &  14.3       &2.50 &  54.8   &  0.193  & 26.9 & 88.1 & 0.566 \\    

 0.829  &  EigenPrism                    & 0.036   & 0.62 &   629      & ---   &   0   &  0.322  & ---    & ---     &  --- \\ 
           &  Est.Equ($W_{\lambda}$)& 0.889   & 2.60 &  6.20       &0.27&  24.1   &  0.061  & 1.62 & 31.6 & 0.097 \\    
   \hline
   \hline
\end{tabular}
%\end{center}
\label{Corr}
\end{table}

\begin{figure}[!h]
\centering
\includegraphics[height=5in,width=6.5in]{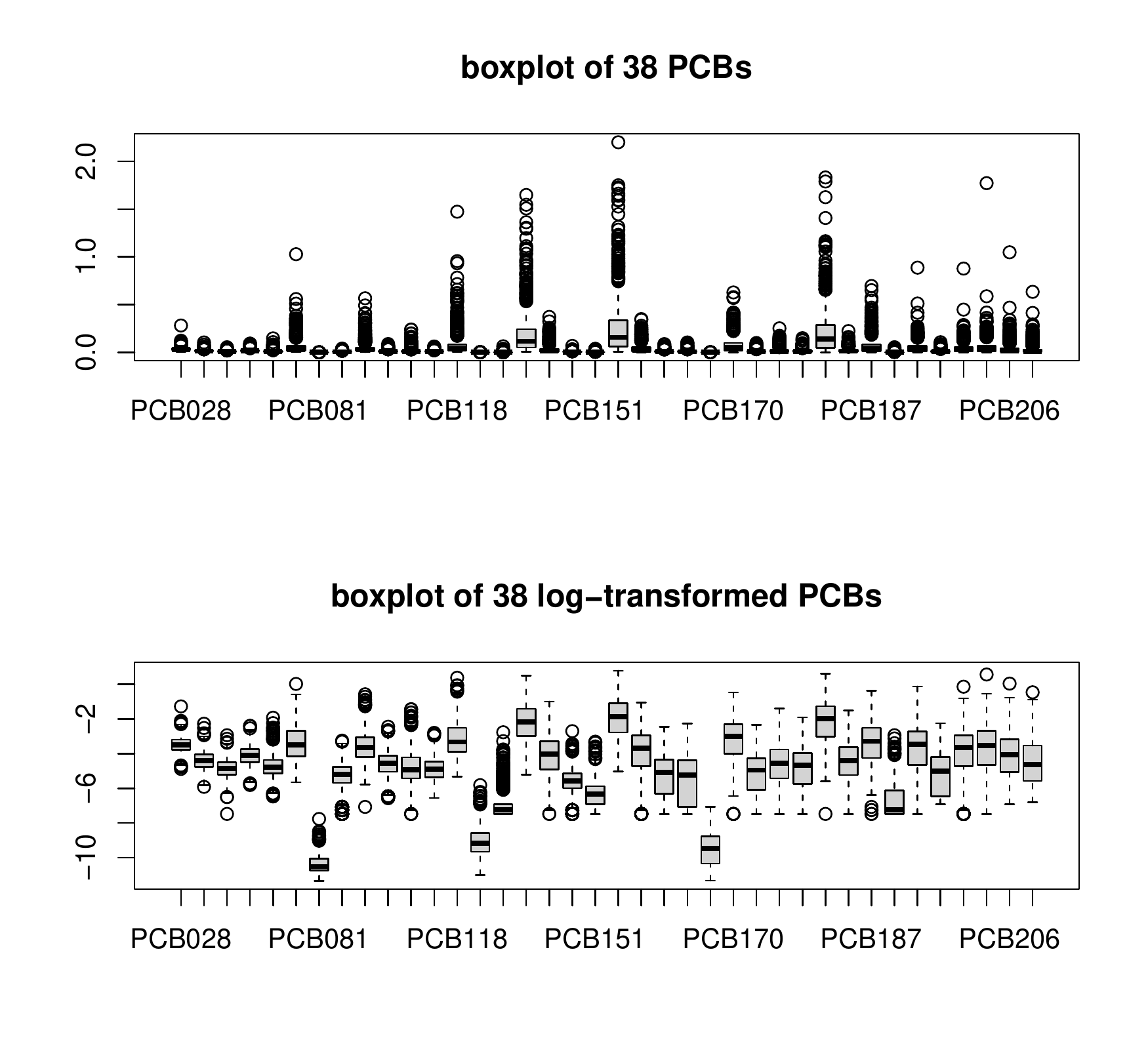}
\caption{Distributions of 38 PCBs before and after logarithm transformation.
 }
\label{boxplot}
\end{figure}

\begin{figure}[!h]
\centering
\includegraphics[height=3in,width=6.5in]{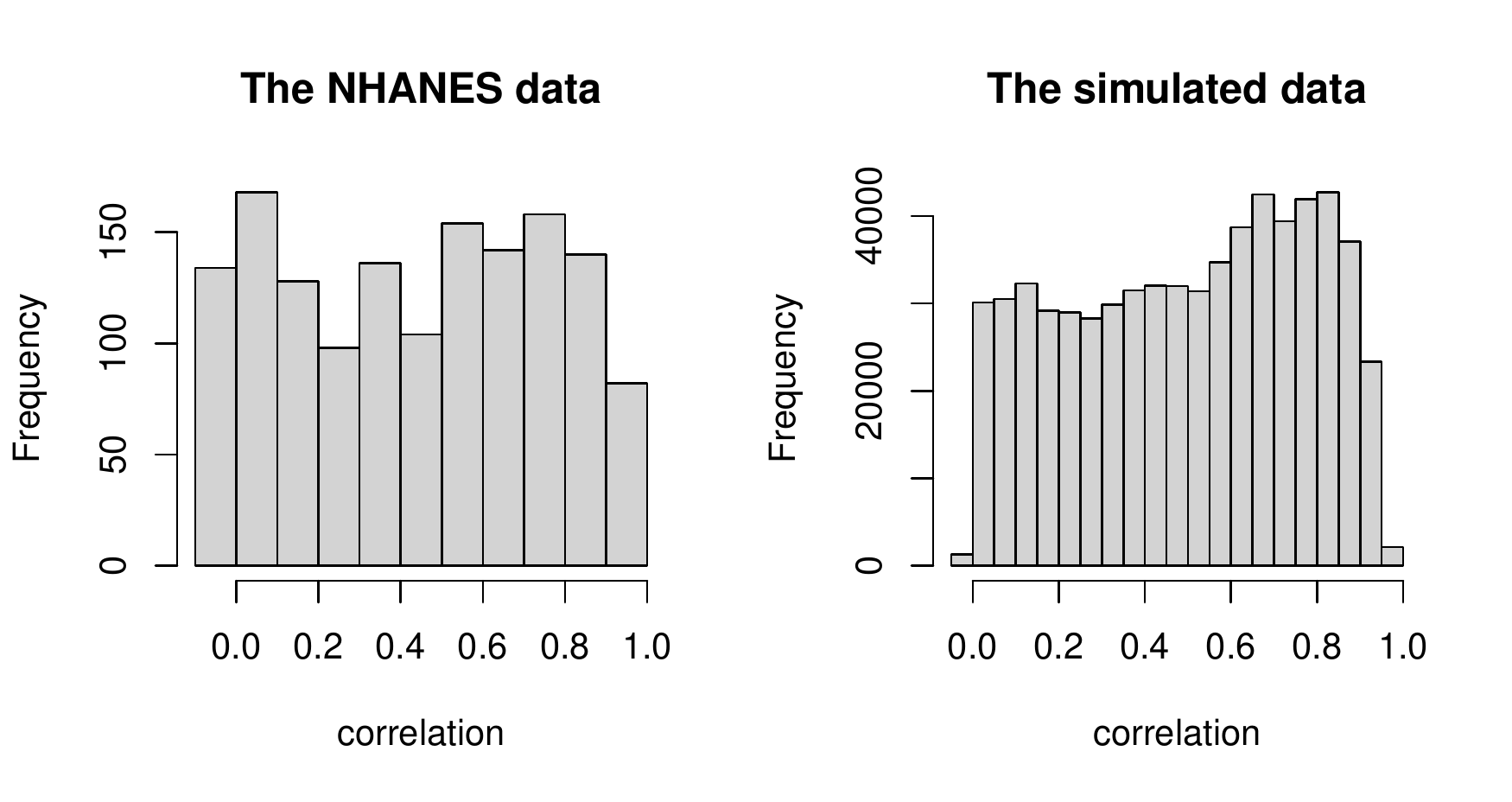}
\caption{
Correlation distributions of 38 PCBs in the NHANES data and the covariates in the simulated data.
 }
\label{corrdistr}
\end{figure}

\begin{table}[!h] 
\caption{Explained variations of glycohemoglobin by the PCB mixture in the NHANES data.}
%\begin{center}
\begin{tabular}{lll ccc} \hline \hline
              &                 &\multicolumn{2}{c}{Assuming Normal}        & \multicolumn{2}{c}{Not assuming Normal} \\
Method        &   $r^{2}$ est.  &  est.var                &  95\% CI        &   est.var            &  95\% CI          \\
\multicolumn{6}{c}{Covariates in the original scale, Main effects} \\
EigenPrism                        &  0.118  &   ---                   & (0.033, 0.193)  &  ---                 & ---  \\
GCTA                                &  0.100  &   $0.81\times 10^{-3}$  & (0.048, 0.159)  &  ---                 & ---  \\
EstEqu($W_{\lambda}$)  &  0.100  &   $1.02\times 10^{-3}$  & (0.038, 0.163)  & $7.27\times 10^{-3}$ & (0.000, 0.267)   \\
TransEE($W_{\lambda}$) &  0.117  &   $0.50\times 10^{-3}$  & (0.074, 0.161)  & $3.55\times 10^{-3}$ & (0.001, 0.234)   \\
EstEqu($W_{*}$)             &  0.119  &   $1.65\times 10^{-3}$  & (0.040, 0.199)  & $13.5\times 10^{-3}$ & (0.000, 0.347)   \\

\multicolumn{6}{c}{Covariates in the original scale, Main and interaction effects} \\
EigenPrism            &  0.401  &   ---                   & (0.276, 0.496)  &  ---                 & ---  \\
GCTA                  &  0.156  &   $2.24\times 10^{-3}$  & (0.066, 0.250)  &  ---                 & ---  \\
EstEqu($W_{\lambda}$) &  0.156  &   $1.22\times 10^{-3}$  & (0.088, 0.225)  & $11.9\times 10^{-3}$ & (0.000, 0.370)   \\
TransEE($W_{\lambda}$) &  0.401  &   $2.73\times 10^{-3}$  & (0.299, 0.504)  & $3.01\times 10^{-3}$ & (0.061, 0.741)   \\
EstEqu($W_{*}$)       &  0.403  &   $3.02\times 10^{-3}$  & (0.295, 0.511)  & $2.29\times 10^{-3}$ & (0.309, 0.497)   \\

\multicolumn{6}{c}{Covariates in the log-transformed scale, Main effects} \\
EigenPrism            &  0.139  &   ---                   & (0.055, 0.212)  &  ---                 & ---  \\
GCTA                  &  0.134  &   $1.22\times 10^{-3}$  & (0.071, 0.207)  &  ---                 & ---  \\
EstEqu($W_{\lambda}$) &  0.134 &   $1.15\times 10^{-3}$  & (0.068, 0.201)  & $9.78\times 10^{-3}$ & (0.000, 0.328)   \\
TransEE($W_{\lambda}$) &  0.128  &   $0.54\times 10^{-3}$  & (0.083, 0.174)  & $4.10\times 10^{-3}$ & (0.003, 0.254)   \\
EstEqu($W_{*}$)       &  0.152  &   $1.53\times 10^{-3}$  & (0.075, 0.229)  & $8.08\times 10^{-3}$ & (0.000, 0.328)   \\

\multicolumn{6}{c}{Covariates in the log-transformed scale, Main and interaction effects} \\
EigenPrism            &  0.272  &   ---                   & (0.120, 0.388)  &  ---                 & ---  \\
GCTA                  &  0.144  &   $1.92\times 10^{-3}$  & (0.056, 0.231)  &  ---                 & ---  \\
EstEqu($W_{\lambda}$) &  0.144  &   $1.31\times 10^{-3}$  & (0.073, 0.215)  & $11.8\times 10^{-3}$ & (0.000, 0.356)   \\
TransEE($W_{\lambda}$) &  0.274  &   $0.38\times 10^{-3}$  & (0.153, 0.396)  & $15.6\times 10^{-3}$ & (0.029, 0.519)   \\
EstEqu($W_{*}$)       &  0.275  &   $4.45\times 10^{-3}$  & (0.145, 0.406)  & $3.37\times 10^{-3}$ & (0.162, 0.389)   \\
\hline\hline
\end{tabular}

The bootstrap sample size for the GCTA approach is $1000$.
%\end{center}
\label{DataAnalysis}
\end{table}